\documentclass{aa}  
\usepackage{natbib}
\setcitestyle{maxnames=3}
\usepackage{graphicx}
\usepackage{afterpage}
\usepackage{subfigure}
\usepackage{stfloats} 
\usepackage{subfloat}
\usepackage{array}
\usepackage{xcolor}
\usepackage{amsmath}  
\usepackage{amsfonts}  
\usepackage{amssymb}  
\usepackage{comment}

\usepackage{booktabs}
\usepackage{tabularx}
\usepackage{multirow}
\usepackage{siunitx}

\usepackage{txfonts}
\usepackage[hidelinks,colorlinks=true,linkcolor=blue,citecolor=blue]{hyperref}

\title{Dynamics and Spectra-Polarimetric Signatures of GRMHD Simulations with Multiple Magnetic Loops}
\author{Raoul Kinadeter \inst{\ref{affil:wuerzburg}}
\and Christian M. Fromm \inst{\ref{affil:wuerzburg},\ref{affil:frankfurt}}
\and Yosuke Mizuno
\inst{\ref{affil:tsungdao},\ref{affil:schoolphy},\ref{affil:keylab},\ref{affil:frankfurt}}
\and Antonios Nathanail 
\inst{\ref{affil:athens}}
\and Matthias Kadler
\inst{\ref{affil:wuerzburg}}
\and Karl Mannheim
\inst{\ref{affil:wuerzburg}}
}
\institute{%
  Institut f\"ur Theoretische Physik und Astrophysik, Universit\"at W\"urzburg, Emil-Fischer-Str. 31, D-97074 W\"urzburg, Germany\label{affil:wuerzburg}
  \and%
  Institut f\"ur Theoretische Physik, Goethe Universit\"at, Max-von-Laue-Str. 1, D-60438 Frankfurt, Germany \label{affil:frankfurt}
  \and%
  Tsung-Dao Lee Institute, Shanghai Jiao Tong University, Shanghai, 201210, People’s Republic of China \label{affil:tsungdao}
  \and %
  School of Physics and Astronomy, Shanghai Jiao Tong University, Shanghai, 200240, People’s Republic of China \label{affil:schoolphy}
  \and %
  Key Laboratory for Particle Astrophysics and Cosmology (MOE) and Shanghai Key Laboratory for Particle Physics and Cosmology, Shanghai Jiao Tong University, Shanghai 200240, People’s Republic of China \label{affil:keylab}
  \and %
  Research Center for Astronomy and Applied Mathematics, Academy of Athens, GR 11527 Athens, Greece \label{affil:athens}
  }
\begin{document}
 
\abstract
{Relativistic jets are a common outcome of accretion onto black holes, yet their presence and variability depend strongly on the magnetic and dynamical state of the accretion flow. While some systems, such as Blazars and Quasars, launch powerful persistent jets, others, including the Galactic Centre black hole Sgr~A$^\star$, show only weak or transient outflows. The physical conditions leading to the onset or suppression of jet activity remain poorly understood.}
{We investigate the accretion flow conditions that produce transient jets or inhibit jet formation, aiming to improve our understanding of black holes that accrete without strong, steady outflows. We further predict observational signatures in total and polarized emission for comparison with recent observations of Sgr~A$^\star$ in the quiescent state from the radio to the $\gamma$-ray regime.}
{We perform three-dimensional GRMHD simulations of an accreting black hole surrounded by a torus threaded by a poloidal multi-loop magnetic field of alternating polarities. We follow the evolution of the accretion rate, magnetic flux, and jet power, and analyze angular momentum transport. In addition, radiative transfer calculations including Compton scattering are used to derive synthetic total and polarized emission.}
{The simulations show strong variability in jet power while the initial magnetic polarity loops accrete, followed by weaker activity at later times as the system approaches a semi-MAD state. The emission from the disk is relatively stable, weakly polarized and consistent with the quiet state values reported for SgrA$^\star$. The resulting jet emission is strongly suppressed, depolarized by Faraday rotation and conversion in the surrounding cold plasma. {Upscattering calculations yield near-infrared (NIR) high energy light curves that respect observational constraints of the quiescent NIR and X-ray fluxes in SgrA$^\star$}}
{}


   \maketitle
   \nolinenumbers
%

\section{Introduction}
\label{sec:introduction}
The supermassive black hole at the center of the Milky Way, Sagittarius~A$^\star$ (Sgr~A$^\star$), is among the best-studied low-luminosity active galactic nuclei.
Precise astrometric measurements of stellar orbits yield a distance of
$d = (8178 \pm 13_{\rm stat} \pm 22_{\rm sys})\,$pc and a black hole mass
$M_{\rm BH} = (4.152 \pm 0.014)\times10^6\,M_\odot$ \citep{Gravity2019}.
Its proximity gives Sgr~A$^\star$ the largest apparent angular size of any known black hole, making it a prime target for very-long-baseline interferometry (VLBI) observations with the Event Horizon Telescope (EHT) at $230\,$GHz \citep{EHTC2022I}.

EHT observations resolve a ring-like emission structure surrounding a central brightness depression consistent with the black hole shadow, with a diameter of $\sim50\,\mu$as and an asymmetric azimuthal brightness distribution.
To interpret these observations, the EHT Collaboration constructed a library of general relativistic magnetohydrodynamic (GRMHD) simulations, followed by general relativistic radiative transfer (GRRT) calculations.
The models span different accretion states (SANE vs.\ MAD), black hole spins, inclinations, and electron thermodynamics, parameterized through the electron heating prescription $R_{\rm high}$.
Applying observational constraints on image morphology, total flux, and variability, early analyses favored a rapidly spinning black hole ($a=0.94$), viewed at low inclination ($i\lesssim30^\circ$), surrounded by a magnetically arrested disk (MAD) \citep{EHTC2022V}.

Spatially resolved polarimetric measurements provide additional constraints.
The emission ring is found to be highly linearly polarized, with an average fractional polarization of $\sim24$--$28\%$ and a coherent counter-clockwise EVPA pattern, while circular polarization reaches peak values of $\sim5$--$10\%$ of the total intensity \citep{EHTC2024VII}.
The unresolved net polarization fractions are $|m_{\rm net}|\sim5\%$ and {$v_{\rm net}\sim-1\%$.}
Incorporating these polarimetric constraints yields a revised best-fit model in which Sgr~A$^\star$ is a rapidly spinning black hole viewed at high inclination ($i\simeq150^\circ$), surrounded by a MAD accretion flow \citep{EHTC2024VIII}.

Sgr~A$^\star$ is strongly variable across the electromagnetic spectrum, from radio to X-rays, on timescales as short as minutes \citep{Witzel2018,Wielgus2022a}.
This variability is particularly pronounced during quasi-simultaneous near-infrared (NIR) and X-ray flares.
Until recently, the mid-infrared (MIR) regime remained observationally unexplored due to limited sensitivity.
New observations with the James Webb Space Telescope (JWST) have detected MIR flares lasting $\sim40$ minutes, accompanied by delayed sub-millimeter counterparts \citep{VonFellenberg2025}.
The steepening of the MIR spectral index during these events suggests synchrotron emission from a radiatively cooling population of electrons accelerated by magnetic reconnection and/or turbulence.

Such flares are commonly associated with compact orbiting hot spots or plasmoids.
Magnetic reconnection in current sheets accelerates electrons into non-thermal power-law tails with Lorentz factors $\gamma\gtrsim10^3$, producing synchrotron emission observable in the NIR.
If a sufficient fraction of electrons is accelerated, inverse Compton scattering of NIR photons can generate corresponding X-ray flares \citep{Yuan2003,Yuan2004,Broderick2005,Ripperda2020,Scepi2022}, although only a subset of NIR flares show X-ray counterparts.
GRAVITY observations of Sgr~A$^\star$ revealed rotating centroid motion and polarization variability during NIR flares, consistent with a compact, polarized hot spot orbiting near the ISCO at low inclination \citep{GravityC2018}.
Similarly, ALMA polarimetric observations during the 2017 EHT campaign detected EVPA rotations and coherent loops in the $QU$ plane following an X-ray flare, interpreted as orbital motion of a hot spot embedded in an ordered magnetic field \citep{Wielgus2022b}.

{In MAD accretion flows, the accumulation of magnetic flux near the black hole can intermittently halt accretion due to the magnetic tension of the accumulated field lines. This can lead to the ejection of hot, under-dense, highly magnetized flux tubes that re-enter and orbit within the disk \citep{Bisnovatyi-Kogan1974, Narayan2003, Porth2021,Najafi-Ziyazi2024}. In contrast the standard and normal evolution (SANE) mode of accretion features a weaker magnetic field and does not oversaturate the black hole with magnetic flux, which leads to a much more steady accretion flow. It has been shown that the accretion modes affect the structure, temperature, magnetization and velocity of the jet, see e.g. \citep{Fromm2022}.}
Such flux eruptions and reconnection events provide a natural framework for hot spot and plasmoid formation in GRMHD simulations \citep{Porth2021,Nathanail2020,Nathanail2022b}.
Multi-loop magnetic field configurations further enhance reconnection and plasmoid production \citep{Nathanail2020,Jiang2023,Chashkina2021,Nathanail2022a,Jiang2024,Jiang2025}.
While reconnection in ideal GRMHD is mediated numerically, studies including explicit resistivity show qualitatively consistent behavior \citep[e.g.,][]{Nathanail2025}.

Despite the lack of jet detection in Sgr~A$^\star$, indirect evidence suggests the presence of weak outflows \citep{Yusef-Zadeh2020}.
Future VLBI facilities such as the ngVLA and ngEHT may enable a definitive detection \citep{Chavez2024}.
Analogies with jet state transitions in X-ray binaries \citep{Fender2004,FenderBelloni2012} and radio-loud AGN \citep{Moravec2022} further motivate studies of jet formation and variability in Sgr~A$^\star$.

In this work, we adopt a multi-loop magnetic field configuration and focus on a single representative model. {The main focus of this paper is to study several diagnostics of the accretion disk dynamics, jet activity, magnetic field properties and angular momentum transport, in order to gain a deeper and more detailed understanding of the dynamics and the evolution of multi-loop models. Secondly, we study signatures of the associated polarized emission at 230\,GHz and compare the results to current observations of the Galactic Center.}
This paper is organized as follows: Section~\ref{sec:numerical_setup} describes the GRMHD and GRRT setup, Section~\ref{sec:results} presents the dynamical and radiative results, and Section~\ref{sec:discussion_conclusion} discusses the implications and conclusions.

\section{Numerical Setup}
\label{sec:numerical_setup}
\subsection{GRMHD Setup}
We performed a {non-axisymmetric } three-dimensional ideal single-fluid GRMHD simulation using the \texttt{KHARMA} code \citep{Prather2024}. We solve the GRMHD equations (see Appendix \ref{sec:GRMHD_appendix}) in the presence of a rapidly spinning ($a = 0.9375$) Kerr black hole in the funky modified Kerr-Schild (FMKS) coordinates $r$, $\theta$, $\phi$ with a resolution of $N_r \times N_\theta \times N_\phi = 192 \times 192 \times 192${, which is sufficient to sustain the MRI-driven turbulence throughout the simulation, \citep{Porth2019, Nathanail2022b} }. The computational domain extends from $0.9r_g$ to $1000r_g$, where $r_g=GM/c^2$ is the gravitational radius. The simulation outputs data every $5\,GM/c^2$, which sets the cadence. Accretion is fed by a torus setup after \cite{FishboneMoncrief1976}. {To initiate accretion from the surrounding torus onto the black hole, the internal energy is perturbed according to $u_g = u_g(1 + X_p)$, where $X_p$ is a random number satisfying $|X_p| < 0.1$. This perturbation seeds the magneto-rotational instability (MRI). As is standard in GRMHD simulations, we impose density and internal energy floors in low-density regions to maintain numerical stability. Specifically, we adopt geometrically scaled floors given by $\rho_{\rm fl}=10^{-6}r^{-3/2}$ for the density and $u_{\rm fl}=10^{-8}r^{-3\hat{\gamma}/2}$ for the internal energy. If the parameters drop below the mentioned floor they are re-set to the floor values.}
The torus contains the multi-loop poloidal magnetic field as an initial condition. The inner edge of the torus is located at $r_\text{in} = 20r_g$, and the pressure maximum is at $r_\text{max} = 40r_g$. We used an adiabatic index of $\gamma=5/3$. The simulation uses geometric units where $M=G=c=1$. The standard initial poloidal single-loop magnetic field configuration is generated from the toroidal vector potential
\begin{align}
\label{vector_potential}
    A_\phi = \text{max}\bigg[\frac{\rho}{\rho_\text{max}}\bigg(\frac{r}{r_{\text{in}}}\bigg)^3 \text{sin}^3\theta \,\, \text{exp}\bigg(-\frac{r}{400}\bigg) - 0.01, 0\bigg],
\end{align}

where $\rho$ is the plasma density $\rho_\text{max}$ is the maximum initial density in the torus, $r$ is the radial coordinate and $\theta$ is the poloidal coordinate, \citep{Gammie2003, Wong2021}.
We multiply $A_\phi$ by $\text{cos}((N-1)\theta) \,\, \text{sin}(2\pi(r-r_\text{in})/\lambda_r)$, where $N$ sets the number of loops and $\lambda_r$ controls the radial loop wavelength, i.e., the distance between consecutive loops. In this paper we use $N=3$ and $\lambda_r = 50 \, r_g$. We evolve the simulation to $30000 \,$M. \cite{Jiang2023} discuss the influence of the loop wavelength $\lambda_r$ on the simulation. They showed that small wavelengths lead to dissipation of magnetic energy already inside the torus at early times, suppressing the development of the MRI and hindering accretion, while in the limit of large values of $\lambda_r$, the simulation produces a typical MAD state. Our choice of $\lambda_r = 50 \, r_g$ is located in a regime where we expect a continuous accretion flow onto the black hole and an accretion state located between the Standard and Normal Evolution (SANE) and MAD state.
{While the use of multiple magnetic loops of alternating polarity may appear artificial, such configurations are in fact expected in realistic accretion flows. First, external feeding mechanisms (e.g. stellar winds; \cite{Ressler2023}) likely supply magnetic flux with stochastic orientations. Second, even in the absence of such variability, MRI-driven turbulence naturally generates a large-scale dynamo that produces cyclic reversals of the magnetic field polarity, as evidenced by the well-known butterfly diagrams in stratified shearing-box and global simulations (e.g. \cite{Brandenburg1995}). Finally, global GRMHD simulations show that magnetic flux is accreted in patches and loops, leading to time-dependent polarity inversions near the black hole (e.g. \cite{Beckwith2008, Liska2018b, Nathanail2022a}). Therefore, alternating-polarity magnetic structures should be regarded as a natural outcome of accretion physics rather than an artificial initial condition.

\subsection{GRRT Setup}
\label{sec:GRRTsetup}
In order to compare our model with observations, we perform general relativistic radiative transfer (GRRT) calculations on the GRMHD simulation data in post-processing. For this, we use the \texttt{IPOLE} code (\cite{Moscibrodzka2018}). \texttt{IPOLE} solves the equations of polarized relativistic radiative transfer along ray-traced null geodesics for a given radiation model. Due to the use of geometric units, the simulation itself is scale-free, meaning there is no reference to physical units in the synthetic data. For comparison with EHT observations, we set the black hole mass $M=4.14\times10^6 \, M_\odot$ and distance $d = 8.14 \,$kpc to that of Sgr~A$^\star$(\cite{Gravity2022}). We iterated $M_{\text{unit}}$ such that the resulting average flux matches the observed EHT value of $\sim 2.4 \,$Jy at $230 \,$GHz (\cite{EHTC2022I}). We list the $M_{\text{unit}}$ values and discuss the conversion to physical units in Appendix \ref{CGSconversion}. We fixed the inclination angle between the jet axis and the observer at $30^\circ$ and used a field of view (FOV) of $150\times150 \, \mu$as with a resolution of $600\times 600$ pixels for the $230 \,$GHz calculations. \\
The dynamics of the plasma in GRMHD simulations are dominated by protons. As such, a simulation only provides information on the proton temperature $T_p$. However, emission and absorption processes in the plasma crucially depend on the electron temperature and on the electron distribution function (eDF) for a given radiation model. In this work, we focus on the effects of thermal synchrotron radiation assuming a Maxwell-J\"uttner eDF. The corresponding thermal emission and absorption coefficients are given in \cite{Pandya2016}. We consider a frequency of $230 \,$GHz for comparison with EHT observations of Sgr~A$^\star$. Regarding the electron temperature, we employ the approach of \cite{Moscibrodzka2016}, the so-called $R-\beta$ model, where

\begin{align}
    \theta_e &= \frac{k_B T_e}{m_e c^2},
    \qquad
    T_e = \frac{2m_p u}{3k_B\rho(2+R)},
    \label{theta_and_T} \\
    R &= \frac{T_p}{T_e}
    = R_{\mathrm{high}}\frac{\beta^2}{1+\beta^2}
    + R_{\mathrm{low}}\frac{1}{1+\beta^2}.
    \label{eq:rhigh}
\end{align}

Here $k_B$ is Boltzmann's constant, $m_p$ and $m_e$ are the proton and electron masses, $u$ and $\rho$ are the internal energy density and rest mass density, respectively, and $\beta = \frac{2P_\text{gas}}{b^2}$ is the plasma-$\beta$ parameter, i.e., the ratio of gas pressure to magnetic pressure. Thus, we can compute the electron temperature in the R-$\beta$ model entirely from plasma quantities in the GRMHD simulation and the two free parameters $R_{\text{low}}$ and $R_{\text{high}}$, which control the relative temperatures between low-$\beta$ (jet) and high-$\beta$ (disk) regions, e.g. \citep{Fromm2022}. In this work, we set $R_{\text{low}} = 1$ and $R_\text{high}=1$ and $R_\text{high}=160$. This choice of $R_{\rm high}$ allows us to explore disk-dominated and jet-dominated emission models.

\begin{figure*}[t]
    \centering
    \includegraphics[width=.85\textwidth]{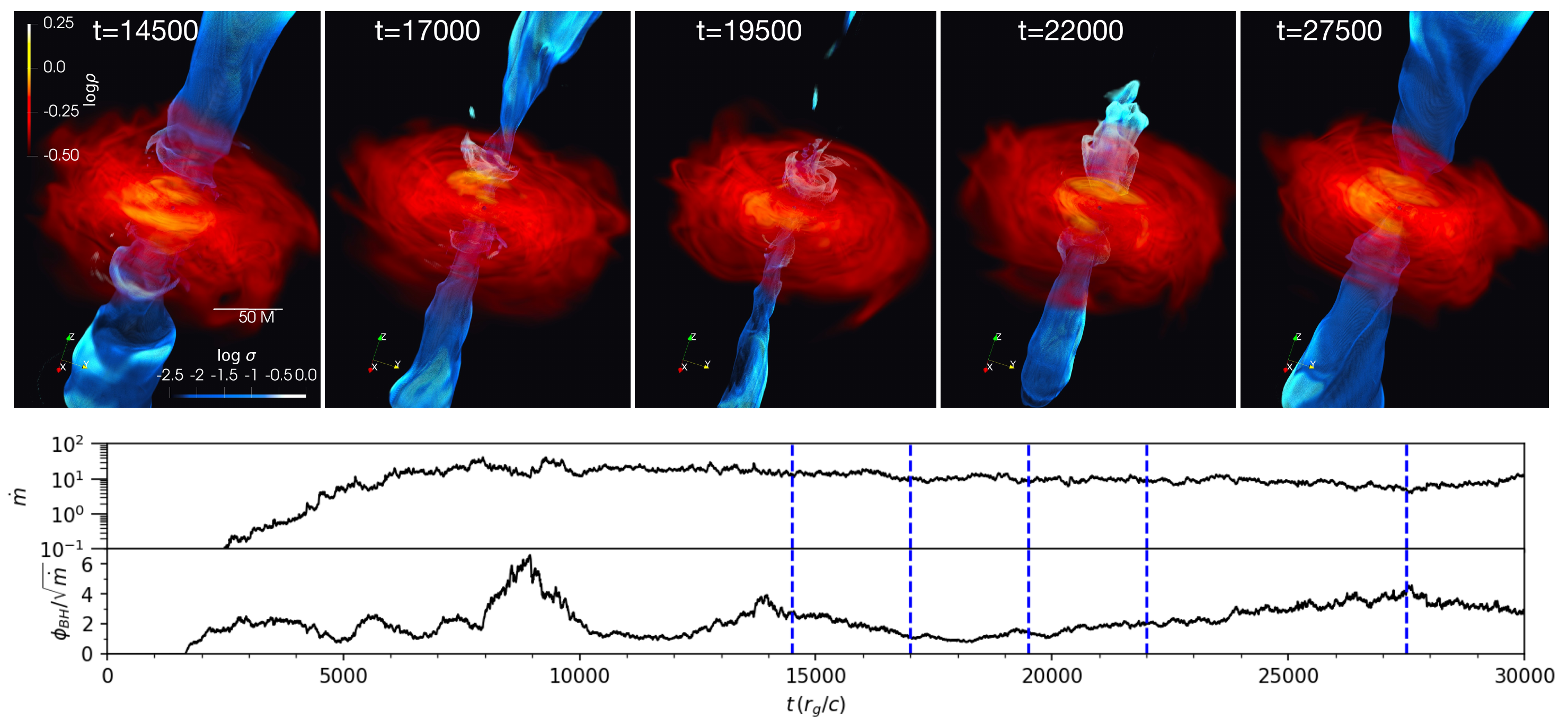}
    \caption{{Top panel:} {3-dimensional rendering of the accretion at different times, indicated by the vertical dashed lines in the figure below. The mass density $\rho$ is colored in red and the magnetization $\sigma$ in blue.} {Bottom panel:} Time series for the mass accretion rate $\dot{M}$ (top), and the MAD parameter $\Phi_{BH}/\sqrt{\dot{M}}$ (bottom); {the model does not reach the MAD regime for which values $\gtrsim 15$ are required.}}
    \label{fig:rendering}
\end{figure*}

\section{Results}
\label{sec:results}
\subsection{GRMHD Simulation}
\label{sec:resultsGRMHD}

\begin{figure*}[h]
    \centering
    \includegraphics[width=.85\textwidth]{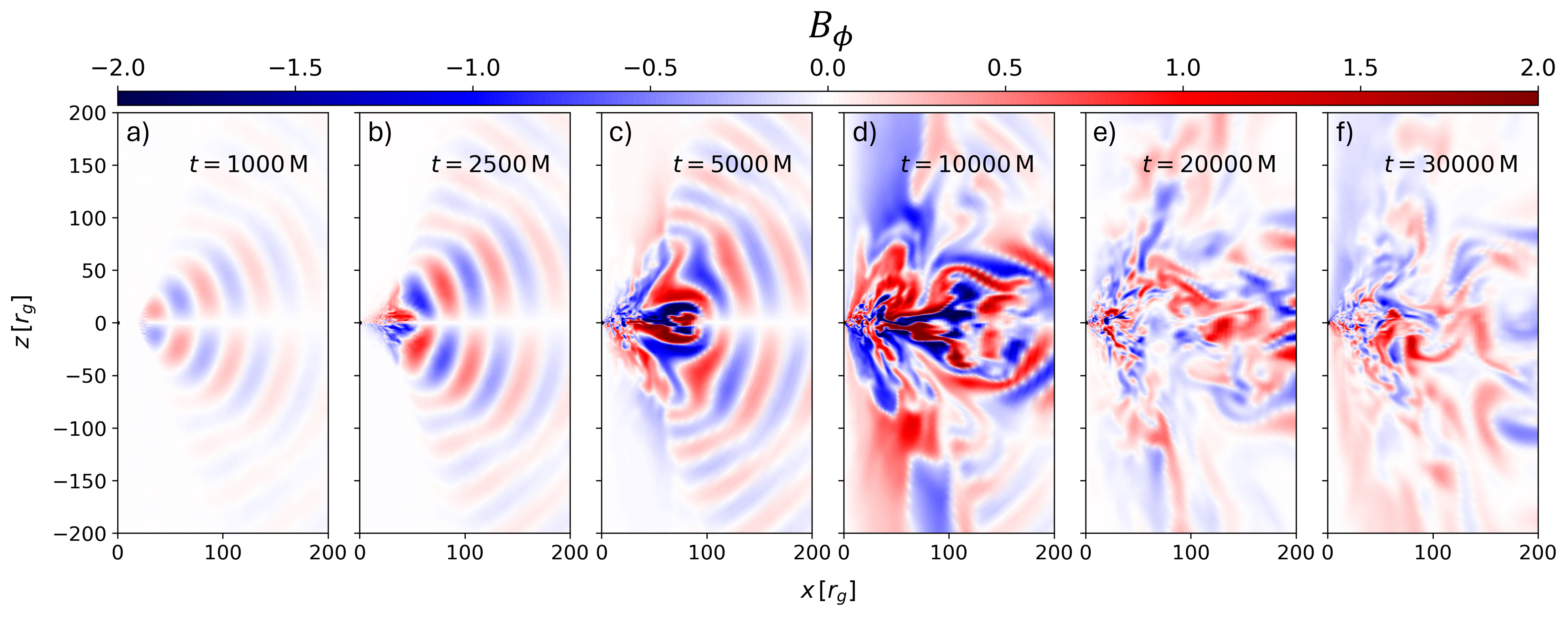}
    \caption{Evolution of the toroidal magnetic field $B_\phi$ in the $x-z-$plane.} 
    \label{fig:bphi_evolution}
\end{figure*}

The results of our simulation are shown in Fig.~\ref{fig:rendering}. {The top image shows a 3-dimensional rendering of the system, where we colored the density $\rho$ in red, which visualizes the torus, and the magnetization $\sigma$ in blue, which makes the jet visible.
We show the mass accretion rate $\dot{M}$ and the MAD parameter $\Phi_\text{BH}/\sqrt{\dot{M}}$ of our multi-loop model on the bottom of Fig. \ref{fig:rendering}.}
They are calculated using the definitions of \cite{Porth2017}: 

\begin{align}
    \label{mdot}
    \dot{M} = \int_0^{2\pi}\int_0^\pi \rho u^r \sqrt{-g} \,d\theta d\phi
\end{align}

\begin{align}
    \label{PhiBH_expr}
        \Phi_\text{BH} = \frac{1}{2} \int_0^{2\pi}\int_0^\pi |B^r| \sqrt{-g} \, d\theta d\phi,
\end{align}
{where $u^r$ and $B^r$ are the velocity and magnetic field strength components, respectively. Both quantities are evaluated as they cross the black hole event horizon.}
Mass accretion saturates at $\sim 6000\,$M and settles into a stable until the end of the simulation from $t\sim 12000\,$M. The magnetic flux $\Phi_\text{BH}$ (see right panel of Fig. \ref{fig:parity}) does not exhibit the periodic dips, which are characteristic of the MAD state, as they signify the expulsion of magnetic flux from the black hole horizon. Instead, there is a slow build-up of magnetic flux, followed by a sharp decrease at $\sim 10000\,$M. This is followed by another phase of flux accumulation and expulsion between $\sim 12000 - 16000\,$M. After that, the flux settles into a stable state. Overall, the behavior of this model is somewhere between SANE and MAD. {After the MRI has fully developed mass accretion rates in SANE models are of the order $\lesssim0.5$ and $\sim5$ in MADs. Typical values for the magnetic flux and MAD parameter are $\Phi_\text{BH}\lesssim1$ and $\Psi \lesssim5$ for SANEs and $\Phi_\text{BH} \sim 25$ and $\Psi \sim 15$ \cite[see e.g.][]{Fromm2022, Dhruv2025}.} The model does not reach the MAD state, for which a MAD parameter of order $\sim 15$ or higher would be required \citep{Tchekhovskoy2011}. This is due to the high accretion rate combined with a rather low magnetic flux. Although the model can reach high absolute values in the magnetic flux, compared to a typical SANE model, it cannot maintain a substantial amount of flux over a long period of time \citep[see also][]{Jiang2023,Jiang2024}.
\\
The standard approach in the literature to explain the behavior of multi-loop models is that magnetic pressure builds up as the first loop is accreted onto the black hole. During this time, a strong jet can form via the Blandford-Znajek mechanism (\cite{BlandfordZnajek1977}).
When the next loop is accreted, it will reconnect with the first loop since it has the opposite polarity, thus causing a breakdown of the magnetic flux and suppression of the jet. The next loop can then be accreted, and the cycle continues. The long-term behavior is then governed by the last loop. 
\\
However, as can be seen in Fig.~\ref{fig:bphi_evolution}, the process does not seem to be that straightforward. The figure illustrates the evolution of the toroidal magnetic field $B_\Phi$. Fig.~\ref{fig:bphi_evolution} a) shows the structure of the toroidal field after $1000\,M$, which is still very close to the initial condition with multiple magnetic field loops of alternating polarity, with first signs of accretion forming on the inner torus edge. Instead of accreting one loop after the other, a kink forms along an axis that is $\sim30^\circ$ to the equatorial plane (Fig.~\ref{fig:bphi_evolution} b). Accretion of magnetic flux occurs along this axis, which leads to a bubble of magnetic field, that was part of the first loop being pushed outward in the equatorial plane. As accretion proceeds, the magnetic field from the next loop flows around the bubble, enveloping it and eventually disconnecting it from the inner accretion flow (Fig.~\ref{fig:bphi_evolution} c). This process grows the bubble and pushes it out further, while the field in the inner torus $\lesssim 50\,r_g$ is completely turbulent and none of the initial field structure is preserved (Fig. \ref{fig:bphi_evolution} d). The structure is reminiscent of a magnetic Rayleigh-Taylor instability combined with MRI, where the lighter magnetic field that has accumulated at the black hole horizon is pushed outward against the dense accretion plasma.
After $\sim 16000\,$M, the field inside the bubble also becomes turbulent, and the structure of the initial ordered loops is completely mixed and destroyed in the inner $\lesssim 200\,r_g$ (Fig. \ref{fig:bphi_evolution} e)-f)). At larger distances, it is still preserved, and it continues to flow around the now turbulent bubble.
{Additionally, we can observe the formation and growth of Kelvin-Helmholtz-like instabilities and mixing of polarities along the jet wall, due to the shearing between the up and outward-moving flux in the jet and the diagonally accreting magnetic field in the sheath layer \citep[see also discussion in ][]{Jiang2023}.}
\\
To understand the dynamics of this particular model in more detail, we study the outflow of energy via the jet power $P_{jet}$ and also calculate the jet efficiency $\eta$ and asymmetry $\mathcal{P}$ \citep[see, e.g.,][]{Tchekhovskoy2011,Nathanail2020}. The jet power compares the rest mass energy of the inflowing matter, as given by the mass accretion rate in Eq. (\ref{mdot}), with the total outflowing energy, using the stress-energy tensor $T^\mu_\nu$:
\begin{align}
\label{Pjet}
    P_{jet} = \int_0^{2\pi}\int_0^\pi (-T^r_t - \rho u^r) \sqrt{-g} d\theta d\phi.
\end{align}
In the integration, we only consider regions of unbounded plasma with Bernoulli parameter $-hu_t > 1.02$ and evaluate the energy flux through a surface at a distance of $r=20 \,r_g$ from the black hole.
The jet efficiency is given by
\begin{align}
\label{Jeteff}
    \eta = {P_{jet}}/{\dot{M}}.
\end{align}
The jet asymmetry is a measure of which jet dominates at a given time. It is given by
\begin{align}
\label{Jetasymm}
    \mathcal{P} =  \left({P_{jet, \text{NH}} - P_{jet, \text{SH}}}\right)/\left({P_{jet, \text{NH}} + P_{jet, \text{SH}}}\right),
\end{align}
where $P_{jet, \text{NH}}$ and $P_{jet, \text{SH}}$ denote the jet power contributions in the northern ($z > 0$) and southern ($z < 0$) hemispheres, respectively. For $\mathcal{P} \simeq -1$ the lower jet dominates, for $\mathcal{P} \simeq 1$ the upper jet dominates, and for $\mathcal{P} \approx 0$ the jet structure is symmetric. 
We plot the above quantities in Fig. \ref{fig:jet_power}. We see the highly variable and intermittent structure in the outflow behavior that matches that of the magnetic flux accumulation (see right panel in Fig. \ref{fig:parity}), which means jet activity is tied to high magnetic flux states. {This alternating presence and absence of the jet is uncommon for typical MAD/SANE GRMHD simulations. These simulations always produce steady jet outflows.} On average, both jets contribute roughly equally to the released power. However, there are also phases where the jet structure becomes significantly asymmetric (see bottom panel of Fig.~\ref{fig:jet_power}).

\begin{figure}[h]
    \centering
    \includegraphics[width=0.5\textwidth]{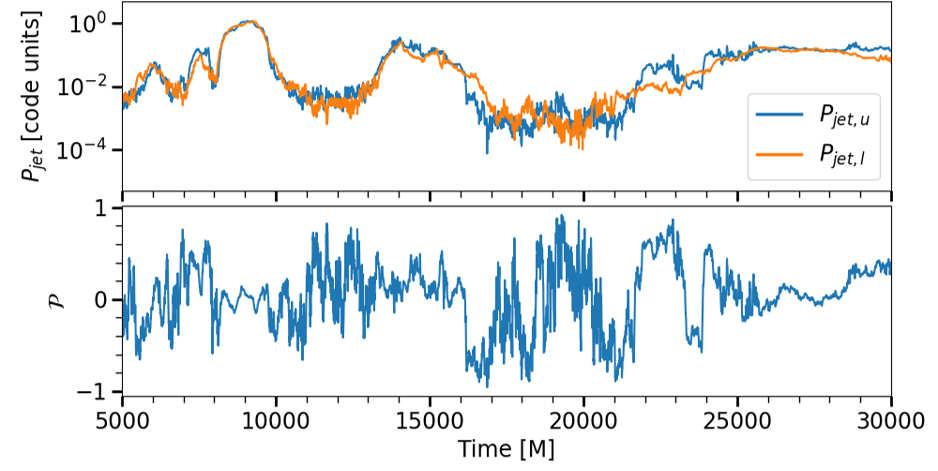}
    \caption{{Top panel:} Jet power of the upper and lower jets, Eq. (\ref{Pjet}). {Bottom panel:} Jet asymmetry measure, as given in Eq. (\ref{Jetasymm}).}
    \label{fig:jet_power}
\end{figure}

As mentioned earlier, the behavior of the jet power and efficiency closely follows that of the magnetic flux profile. To further investigate this correlation and explore possible explanations, we study changes in the topology of the magnetic field. Therefore, we calculate the parity $C$ between the dipole-like $U^D$ and quadrupole-like $U^Q$ magnetic field components \citep{Flock2012,Kaaz2025}, which is defined similarly to the jet asymmetry in Eq. (\ref{Jetasymm})
\begin{align}
\label{parity}
    C(U^D, U^Q) =  \frac{U^D - U^Q}{U^D + U^Q},
\end{align}
where the components are given by
\begin{equation}
\label{dipole_quadrupole_components}
\begin{aligned}
    U^D = (B_r^{\text{AS}})^2 + (B_\theta^{\text{S}})^2 + (B_\phi^{\text{AS}})^2 \\
    U^Q = (B_r^{\text{S}})^2 + (B_\theta^{\text{AS}})^2 + (B_\phi^{\text{S}})^2.
\end{aligned}
\end{equation}
The symmetric and antisymmetric components above are defined for $i = r, \theta, \varphi$ with
\begin{equation}
\label{anti_symm_components}
\begin{aligned}
    B_i^{S} = \frac{1}{2}(B_i^{\text{NH}} + B_i^{\text{SH}}) \quad
    B_i^{AS} = \frac{1}{2}(B_i^{\text{NH}} - B_i^{\text{SH}})
\end{aligned}
\end{equation}
and $B_i^{NH}$ and $B_i^{SH}$ are the volume-weighted averages over a spherical shell of the respective magnetic field components:
\begin{align}
\label{NH_SH_average}
    B_i^{\text{NH/SH}} = \frac{\int_0^{\pi}\int_0^{2\pi} B_i \sqrt{-g} \, d\theta \,d\varphi} {\int_0^{\pi}\int_0^{2\pi} \sqrt{-g} \, d\theta \,d\varphi},
\end{align}
where the integrand is set to zero according to the considered hemisphere. When the parity $C \simeq 1$, the dipole-like (odd) component of the magnetic field dominates, whereas for $C \simeq -1$ it is the other way around, and the magnetic field topology is rather quadrupole-like (even). When $C\simeq0$, neither component dominates. In Fig. \ref{fig:parity} we show a color map of the parity parameter as a function of time and distances $r \lesssim 450\,r_g$, along with the temporal evolution of the magnetic flux $\phi_{BH}$ and the jet efficiency $\eta$ for comparison. We observe that the field topology is mostly dipole-like (red) during phases of increasing and high states of magnetic flux with occasional interruptions during flux eruption events, where it transitions to a mixed (yellow) or slightly quadrupole-like (green-blue) state. 
\begin{figure}[h!]
    \centering
    \includegraphics[width=0.5\textwidth]{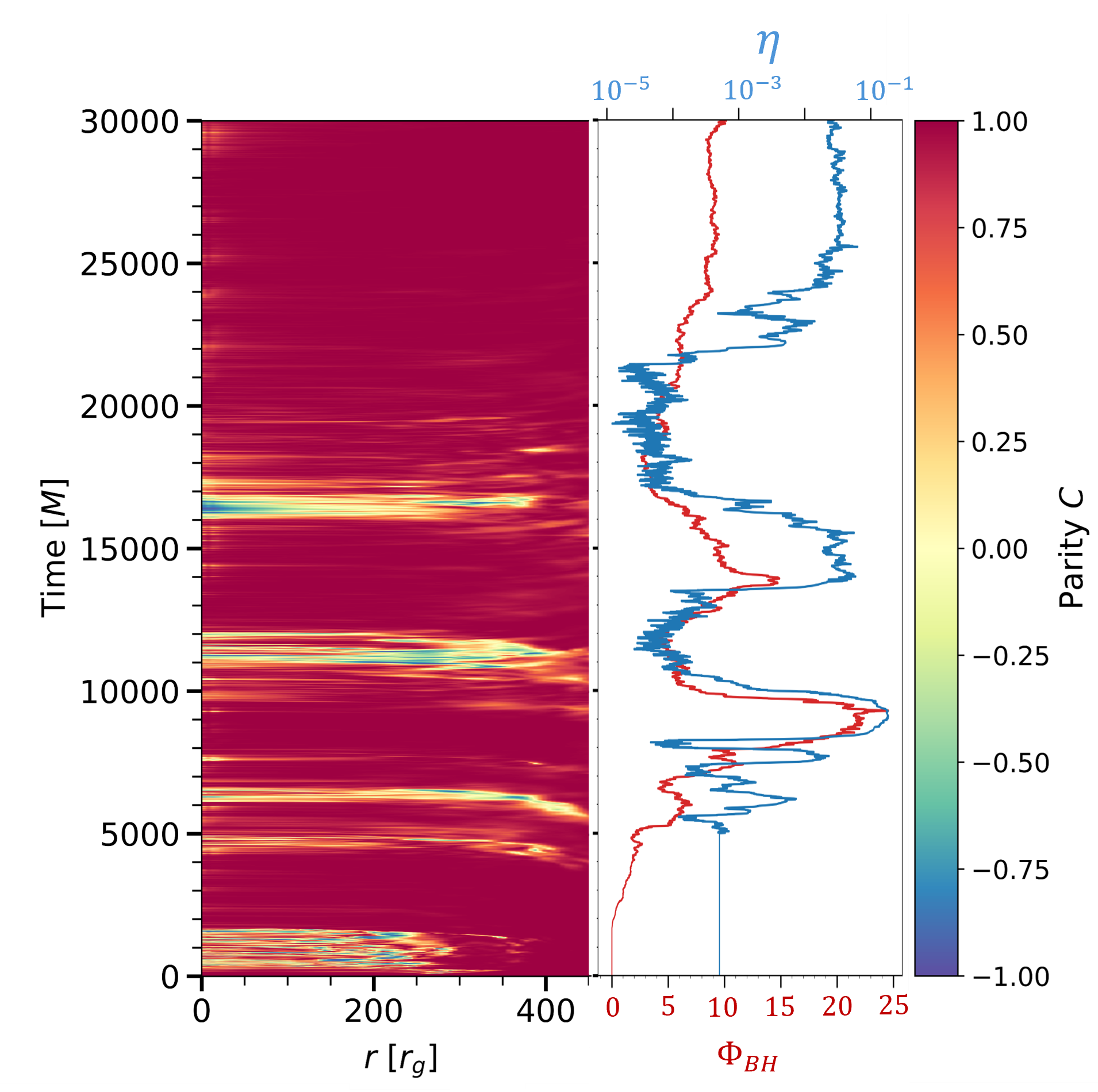}
    \caption{{Left:} Spacetime diagram of the magnetic field parity $C$ (see Eq. (\ref{parity})) on a spacetime plot. {Right:} Magnetic flux $\phi_{BH}$ (red) and jet efficiency $\eta$ (blue) for comparison of the temporal evolution.}
    \label{fig:parity}
\end{figure}
This transition is clearly dependent on the distance from the black hole. Small dips in $\phi_{BH}$ only affect the parity $C$ in a region very close to the black hole, which can be seen in the time interval $t \sim 20000-30000\,$M, where the overall flux trend increases with occasional small dips. In contrast, decreases that make $\phi_{BH}$ fall below a certain threshold can affect the field topology across larger scales, e.g., $t \sim 11000\,$M and $\sim 16000\,$M.
We also note that the field right at the beginning of the simulation is dipole-like at all distances but becomes mixed/quadrupole-like only at distances $r \lesssim 300\,r_g$ during the early accretion phase between $t \sim 300 - 1700\,$M. At later times, this transition region expands to $r \gtrsim 400\,r_g$. In parts, we can connect this to the evolution of the toroidal field $B_\phi$, as discussed in Fig.~\ref{fig:bphi_evolution}. At the beginning of the simulation, the initial field configuration starts to get warped by the onset of accretion, stronger at small distances and only slightly at larger distances. {This produces a quadrupole in the $B_\phi$ structure with accretion along the kinks.} At later times, accretion has mixed the initial field also at larger scales.
A comparison with the jet power in Fig.~\ref{fig:jet_power} and the efficiency $\eta$ hints that the temporal evolution of the jet activity, especially the intermittent behavior of the re-/disappearing jet, is connected to transitions of the magnetic field topology.
\\
We calculated the density-weighted velocity profile as a function of the radius of the accreting fluid, using
\begin{align}
    \label{mdot}
    \langle \Omega \rangle_\rho (r, t) = \frac{\int_0^{2\pi}\int_0^\pi \rho \, \Omega \,  \sqrt{-g} \,d\theta d\phi}{\int_0^{2\pi}\int_0^\pi \rho \, \sqrt{-g} \,d\theta d\phi},
\end{align}
where $\Omega = u^\phi/u^t$ is the angular velocity of the plasma, and the time average was taken in the time interval $t=5000$-$30000\,$M.  At distances up to $r \sim 100 \,$M, the velocity profile of the multi-loop model closely matches the ideal Keplerian rotation profile, and at larger distances, the motion becomes sub-Keplerian. In this regard, our multi-loop model shows the same behavior as the SANE model with spin parameter $a=0.9375$ studied by \cite{Porth2021}.

\begin{figure*}[t!]
    \centering
    \includegraphics[width=0.85\textwidth]{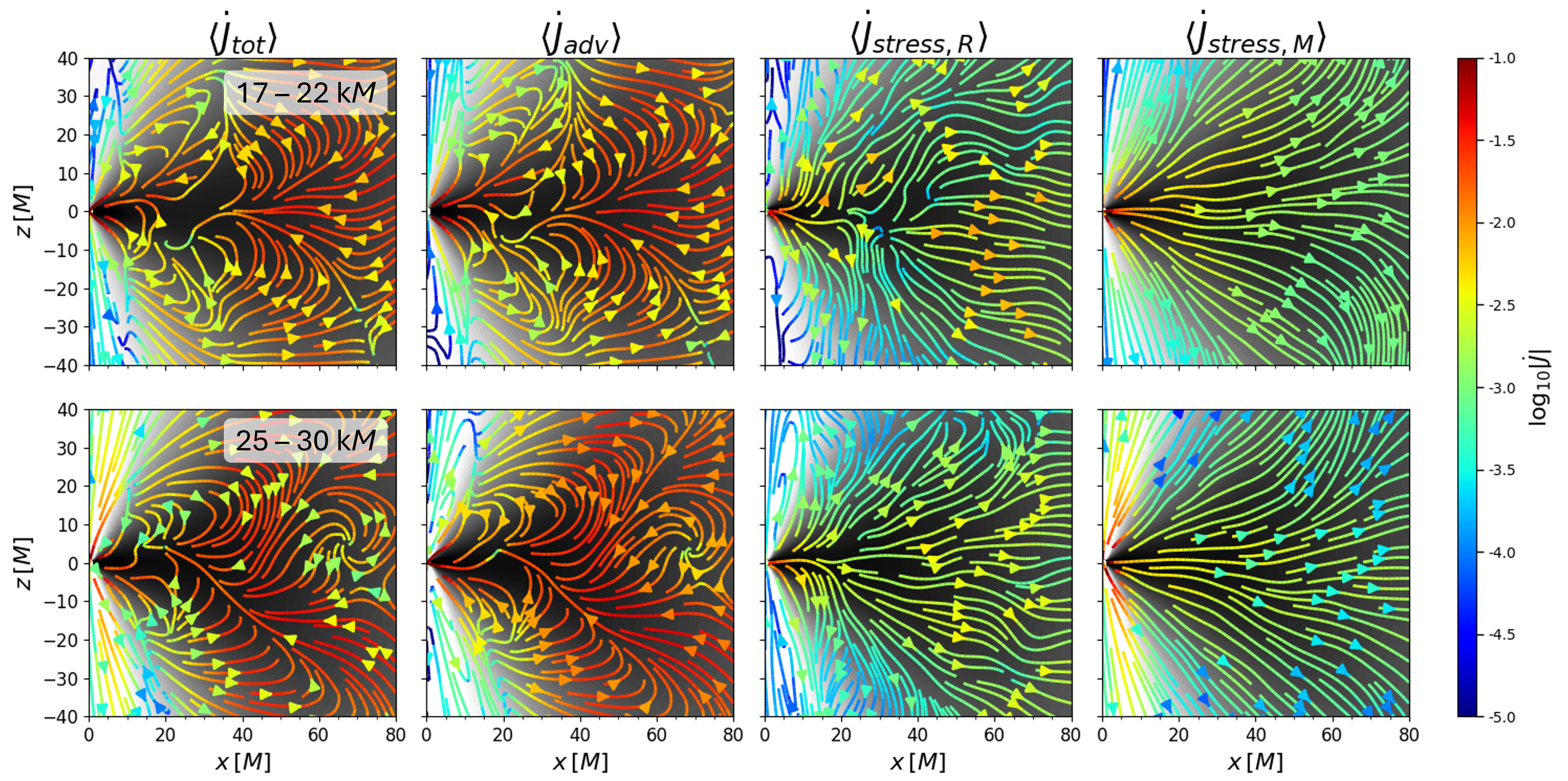}
    \caption{Stream lines of the angular momentum fluxes $\langle \dot{J}_{tot} \rangle$, $\langle \dot{J}_{adv} \rangle$, $\langle \dot{J}_{stress, R} \rangle$ and $\langle \dot{J}_{stress, M} \rangle$, averaged over the time intervals from $17-22\,\text{k}M$ (top panels) and $25-30\,\text{k}M$ (bottom panels) with low and high jet efficiency $\eta$, respectively, see right panel of Fig. \ref{fig:parity}. The color map indicates the absolute magnitude of the respective flux. In the background, we show the density distribution in grey.}
    \label{fig:angularMomentum}
\end{figure*}

\subsubsection{Angular Momentum Transport}

We deepen the analysis of the connection between the jet and the magnetic field in the accretion flow by studying the angular momentum transport in the accretion flow, following \citet{Chatterjee2022}. The total angular momentum component $i=r, \theta$ is given by averaging the respective components of the stress-energy tensor $T^i_\varphi$ over the azimuthal angle $\varphi$ and time $t$:
\begin{align}
    \label{j_tot}
    \dot{J}^i_\text{tot}(r, \theta) = \langle T^i_\varphi \rangle_{\varphi, t} \,,
\end{align}
where $T^i_\varphi$ is given by
\begin{align}
    \label{energy_momentum_tensor}
    T^i_\varphi = (\rho + \gamma_\text{ad}u_g + b^2)u^iu_\varphi - b^ib_\varphi \, .
\end{align}
The total angular momentum flux $\dot{J}_\text{tot}$ can be split into an inward-directed advective term and an outward-directed stress-induced term, \cite{Penna2010}. The advective term is given by
\begin{align}
    \label{j_adv}
    \dot{J}^i_\text{adv}(r, \theta) = \bigg\langle \bigg(\rho + u_g + \frac{b^2}{2} \bigg) u^i\bigg\rangle_{\varphi, t} \langle u_\varphi \rangle_{\varphi, t} \, .
\end{align}
Finally, the outgoing term is defined as the residual from the difference between the total and advective flux
\begin{align}
    \label{j_stress}
    \dot{J}_\text{stress} = \dot{J}_\text{tot} - \dot{J}_\text{adv} \, .
\end{align}
The ingoing, advected flux in Eq. (\ref{j_adv}) contains the product of the mean velocity fields $\langle u^r \rangle \langle u_\varphi\rangle$, while the correlated fluctuations $\langle u^r u_\varphi \rangle$, corresponding to the Reynolds stress are part of the outgoing angular momentum transport \citep{Penna2010}. The time averaging of the above quantities was performed on two different time intervals of our simulation. For the first interval, we chose $t=17-22\,\text{k}M$, where the jet efficiency $\eta$ takes small values. In the second interval from $t=25-30\,\text{k}M$, the jet efficiency is high, as seen in the right panel of Fig. \ref{fig:parity}.
The outgoing viscous angular momentum flux $\dot{J}_\text{stress}$ can be further decomposed into contributions from the Maxwell and Reynolds stresses. The Maxwell stress is given by
\begin{align}
    \label{j_stress,M}
    \dot{J}^i_\text{stress, M}(r, \theta) = \bigg\langle \frac{b^2}{2}u^iu_\varphi - b^ib_\varphi \bigg\rangle_{\varphi, t} \, .
\end{align}
The expression for the Reynolds stress reads
\begin{align}
    \label{j_stress,R}
    \dot{J}^i_\text{stress, R}(r, \theta) = \bigg\langle \bigg( \rho + u_g + \frac{b^2}{2} \bigg) u^i u_\varphi \bigg\rangle_{\varphi, t} - \dot{J}^i_\text{adv} \, .
\end{align}
We show the streamlines of the different angular momentum fluxes from Eqs. (\ref{j_tot}), (\ref{j_stress,M}), (\ref{j_adv}) in the two time intervals with low (top panel) and high (bottom panel) jet efficiency $\eta$ overlaid on the density distribution, averaged over the same time intervals, in Fig. \ref{fig:angularMomentum}. To properly plot the streamlines, we map the coordinate grid from the simulation onto a Cartesian grid and then transform the $r$ and $\theta$ components of the different fluxes to Cartesian $x$ and $z$ vector components, which can be plotted on the uniform grid. In contrast to the treatment in \cite{Chatterjee2022}, we did not symmetrize the fluxes across the midplane.
\\
We begin the discussion of Fig. \ref{fig:angularMomentum} by first focusing on the top row in the interval from $t=17-22\,\text{k}M$ with low jet efficiency. We observe that the turbulent behavior of the advective flux $\langle \dot{J}_\text{adv} \rangle$ also dominates the disk region of the total flux $\langle \dot{J}_\text{tot} \rangle$. In the jet region, there is no outward flux in $\langle \dot{J}_\text{tot} \rangle$, meaning there can be no extraction of energy from the black hole, and thus no jet activity, as we already saw in our discussion of the jet power above; see Fig. \ref{fig:parity}. The Maxwell component $\langle \dot{J}_\text{stress, M} \rangle$ flows outward everywhere; however, its magnitude is small and not able to produce significant outflow in $\langle \dot{J}_\text{tot} \rangle$. 
When we turn our attention to the second row and the interval from $t=25-30\,\text{k}M$ with high jet efficiency, we observe some similar behaviors as in the previous case. The disk region of the total flux is still dominated by the turbulence of the advective component. However, there is also outward flux in the jet region, even though relatively small in magnitude. The Maxwell stress $\langle \dot{J}_\text{stress, M} \rangle$ is still outflowing everywhere, but the magnitude of the outflow in the jet region is several magnitudes higher than before. It leads, in turn, to net outward flowing total angular momentum $\langle \dot{J}_\text{tot} \rangle$ with comparatively high magnitudes. This means that in this case, energy can be extracted from the black hole by the production of a jet, which we also saw earlier.

\subsection{Polarized Radiative Transfer}
In this section, we will present the results of our GRRT calculations, including unresolved, i.e., image-integrated, light curves as well as linear and circular polarization fractions. {We focus on models with $R_{\rm high}=1, 160$, which allows us to study both disk-dominated and jet-dominated emission. Notably, jet-dominated emission ($R_{\rm high}=160$) is part of the best-bet model by the EHT for both SANE and MAD accretion modes, \cite{EHTC2022V}.} In addition, we compute high-resolution images with over-plotted electric vector position angles (EVPAs) and perform a Fourier decomposition of the polarization.
We perform radiative transfer calculations for the last 20\,kM which ensures a stable steady mass accretion rate (see bottom panel in Fig. \ref{fig:rendering}).

\subsubsection{230\,GHz light curves and polarization fractions}
\label{sec:230GHzLightCurvesAndPolarizationFractions}
Fig.~\ref{fig:pol_frac} presents the 230\,GHz light curves for the simulations with 
$R_{\rm high}=1$ (top panel) and $R_{\rm high}=160$ (bottom panel). 
The $R_{\rm high}=1$ model exhibits a relatively steady flux density of around 2\,Jy, 
with only modest short-timescale fluctuations, yielding a mean flux of 2.4\,Jy 
over the analyzed interval (consistent with the flux normalization described in 
Sect.~\ref{sec:GRRTsetup}).
 \begin{figure}[h!]
    \centering
    \includegraphics[width=0.45\textwidth]{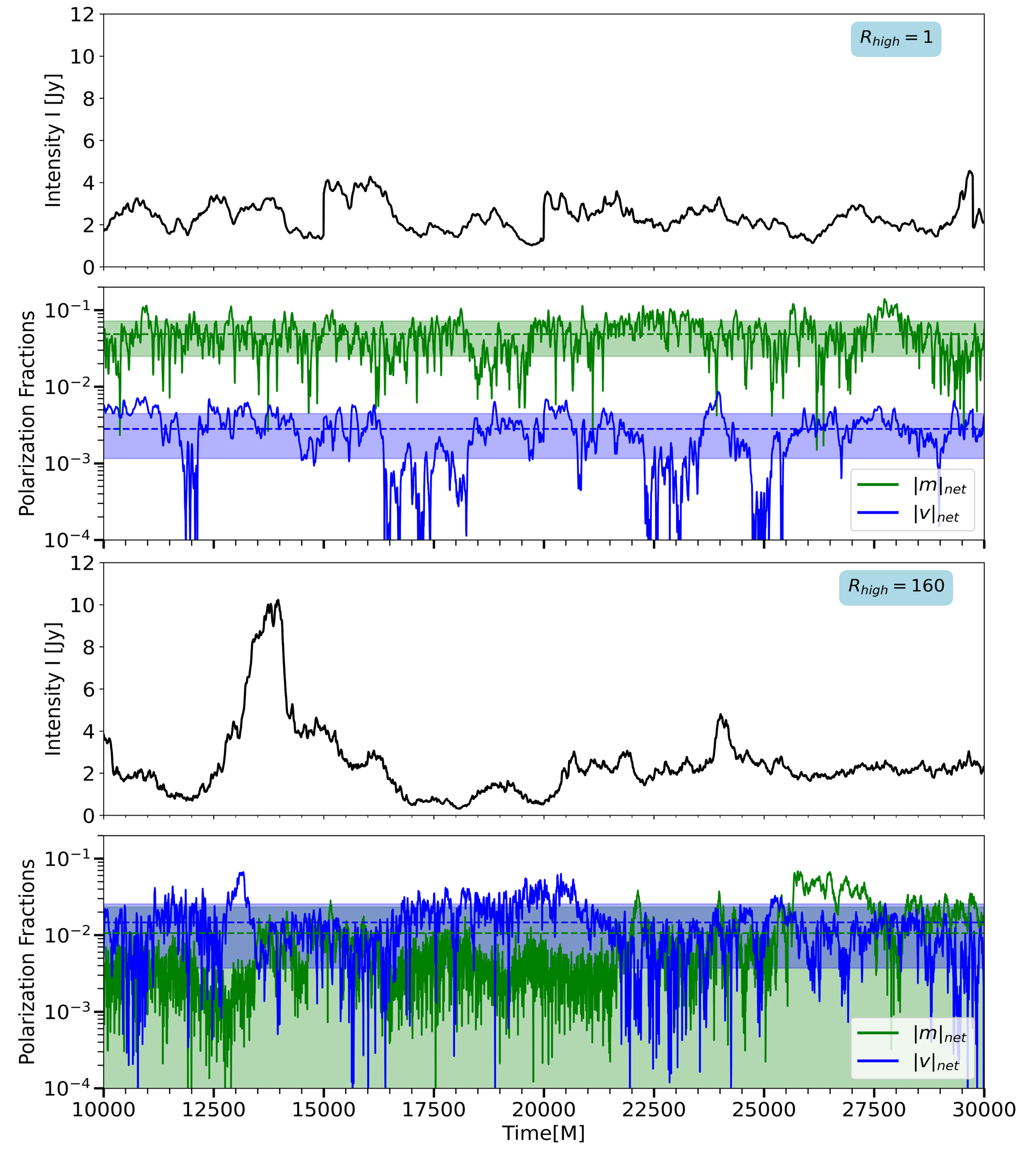}
    \caption{{Top panels:} Light curves at $230 \,$GHz {Bottom panels:} Unresolved linear and circular polarization fractions $m_{net}$ and $v_{net}$, respectively, for different values of R$_{\text{high}}$ at $230 \,$GHz. The green/blue dashed lines and shaded regions indicate the respective mean values and standard deviation.}
    \label{fig:pol_frac}
\end{figure}
In contrast, the $R_{\rm high}=160$ light curve displays 
pronounced variability: a strong flare peaks at $\sim 10$\,Jy near 
$t\approx 13500\,M$, followed by a gradual decline over the subsequent 
$\sim 7000\,M$. At later times ($\gtrsim 20\,kM$), the flux 
settles to a baseline level of $\sim 2$\,Jy, punctuated by moderate flares around 
$t\approx 21\,kM$ and $23\,kM$. As imposed by our flux scaling, 
the time-averaged flux density over the interval $10\!-\!30$\,kM is 2.4\,Jy for both models.

\begin{figure*}[b!]
    \centering
    \includegraphics[width=0.8\textwidth]{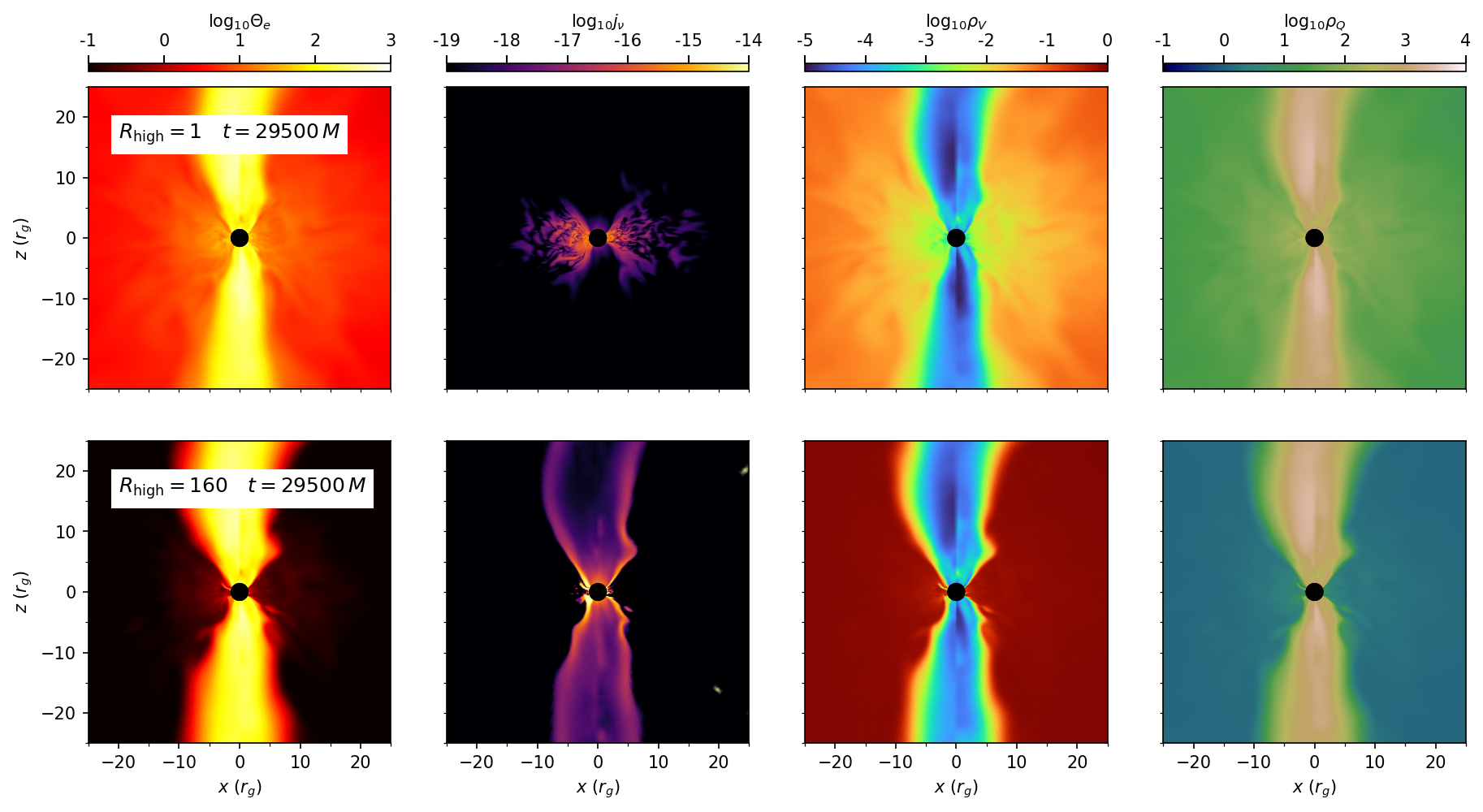}
    \caption{Electron temperatures, $\theta_e$, thermal emissivities at 230\,GHz, $j_{\nu}$, as well as the rotativities $\rho_Q$, and $\rho_V$ at $R_\text{high}=1$ (top) and $R_\text{high}=160$ (bottom) at t=29.5\,kM. }
    \label{fig:electron_temp_rotativities}
\end{figure*}

The corresponding unresolved (image-integrated) polarization fractions are shown in the 
second and fourth panels of Fig.~\ref{fig:pol_frac}. We report the net linear and circular polarization 
fraction 
\begin{align}
    m_{\rm net}=\frac{\sqrt{(\sum_i Q_i)^2+(\sum_i U_i)^2}}{\sum_i I_i},\quad v_{\rm net}=\frac{\sum_i V_i}{\sum_i I_i},
\end{align}
where $(I_i,Q_i,U_i,V_i)$ denote the Stokes parameters at each image pixel. For the 
$R_{\rm high}=1$ model, the linear polarization fraction fluctuates around a mean value 
of $\sim4.8\pm2.4\%$, while the circular polarization fraction remains lower, with a mean of 
$\sim0.3\pm0.2\%$. Both quantities show rapid variability that broadly tracks changes in the 
total intensity. The behaviour differs noticeably in the $R_{\rm high}=160$ model: 
here, the circular polarization fraction exceeds the linear polarization fraction over 
extended periods of the evolution. For this model, both $m_{\rm net}$ and $v_{\rm net}$ 
are around $0.1\pm0.1\%$, and the large error indicates the strong variability in both quantities, which do not closely 
follow the structure of the total-intensity light curve. Note that polarimetric constraints of Sgr~A$^\star$ at $230\,$GHz are $m_{net}\sim7\%$ and $v_{net} \le 1\%$ \citep{Johnson2015,Bower2018,Wielgus2022b,EHTC2024VIII}, which is consistent with our model only for $R_\text{high}=1$.\\
The total and polarization behavior of our models can be understood by considering the nature of the emissivities together with the Faraday rotation and conversion coefficients for a thermal electron distribution. We first turn to the thermal synchrotron emissivities in Stokes $I$, $Q$, and $V$. Following \citet{Dexter2016}, they can be written as:
\begin{equation}
j_{I,Q}=\frac{n e^2 \nu}{2\sqrt{3}c \theta_e^{2}}I_{I,Q}(x),\quad
j_V=\frac{2 n e^2 \nu \cot\theta_B}{3\sqrt{3} c \theta_e^{3}} I_V(x),
\label{eq:themiss}
\end{equation}
where $n$ is the electron density and dimensionless frequency $x\equiv \frac{\nu}{\nu_c}$, with critical frequency $\nu_c=\frac{3}{2}\nu_B \sin\theta_B \theta_e^2$ and electron cyclotron frequency $\nu_B=eB/(2\pi m_e)$. Thus, the dimensionless frequency, dropping the pitch angle dependency for clarity, can be written as
$x\propto \nu/(B\,\Theta_e^{2})$.
The approximations to the synchrotron integrals \citep[see, e.g][]{Dexter2016} can be divided depending on $x$ in an exponentially suppressed $(x\gg 1\,\mathrm{for}\,\theta_e\ll1   \rightarrow {\rm cold\, regime})$ and a power-law regime $(x\ll1\,\mathrm{for}\,\theta_e\gg 1 \rightarrow {\rm hot\,regime})$ 
\begin{eqnarray}
&I_{I,Q,V}(x)&\propto \exp\left(-x^{1/3}\right) \quad (x\gg1\,\mathrm{cold\, regime})\\
&I_{I,Q}&\propto x^{-2/3},\quad I_V\propto x^{-1} \quad (x\ll1\,\mathrm{hot\,regime}).
\end{eqnarray}
Substituting the expressions above and the definition of $x$ into Eq. \ref{eq:themiss} we obtain for the exponentially suppressed regime, i.e. $x\gg1$ (cold regime):
\begin{eqnarray}    
j_{I,Q} \propto n\theta_e^{-2}\exp\left[-\left(B\theta_e^2\right)^{-\frac{1}{3}}\right]\quad
j_V \propto n\theta_e^{-3}\exp\left[-\left(B\theta_e^2\right)^{-\frac{1}{3}}\right],
\label{eq:emiscold}
\end{eqnarray}
and for the power-law regime ($x\ll1$, hot regime):
\begin{equation}
j_{I,Q} \propto n\,B^{2/3}\theta_e^{-2/3},\quad
j_V \propto nB\theta_e^{-1}
\label{eq:emishot}
\end{equation}
After obtaining the behavior of the emissivities in the cold and hot regimes, we perform the same analysis for the thermal rotativities. Faraday rotation, quantified by $\rho_V$, rotates the plane of linear polarization during propagation and reduces the observed linear polarization fraction. Faraday conversion, quantified by $\rho_Q$, converts linear into circular polarization. Their relative strength therefore determines the observed balance between $m_{\rm net}$ and $v_{\rm net}$. 

Following \citet{Shcherbakov2008,Dexter2016}, the coefficients for a relativistic Maxwellian, keeping only the terms including the dimensionless electron temperature $\theta_e$ can be written as:
\begin{align}
    \rho_V \propto \frac{K_0(1/\theta_e)}{K_2(1/\theta_e)} \qquad
    \rho_Q \propto  \frac{K_1(1/\theta_e)}{K_2(1/\theta_e)} + 6 \theta_e ,
    \label{eq:rhoQ_DexterB4}
\end{align}
where $K_n$ are modified Bessel functions. In the ultra-relativistic limit ($\theta_e \gg 1$), the Bessel ratios behave as 
$K_0/K_2 \sim 1/(2\theta_e^2)$ and $K_1/K_2 \sim 1/(2\theta_e)$ which simplifies the equations above to: 
\begin{align}
    \rho_V \propto \theta_e^{-2} \qquad
    \rho_Q \propto \theta_e.
    \label{eq:rothigh}
\end{align}
These simplifications apply mainly to the jet spine and sheath, given the high electron temperatures (independent of the choice of $R_{\rm high}$). However, the wind and disk regions are highly affected by the choice of $R_{\rm high}$. Therefore, we turn next to the cold, sub-relativistic regime, i.e. $\theta_e<<1$. In this case, the ratio of the Bessel functions, i.e., the rotativities, can be written as:
\begin{align}
    \rho_V \propto 1-2\theta_e\qquad
    \rho_Q \propto 1+43/8\theta_e.
    \label{eq:rotlow}
\end{align}
As expected for cold electrons both rotativities approach 1. In Fig. \ref{fig:electron_temp_rotativities} we compute the electron temperature $\theta_e$, the Stokes I emissivity at 230\,GHz, $j_{\nu}$, and rotativities $\rho_{V,Q}$ for $R_{\rm high}=1$ (top) and $R_{\rm high}=160$ (bottom). Notice, that we used the M$_{\rm unit}$ reported in Sect. \ref{sec:GRRTsetup} in the calculation of the emissivities. \\

\begin{figure*}[t]
    \centering
    \includegraphics[width=\textwidth]{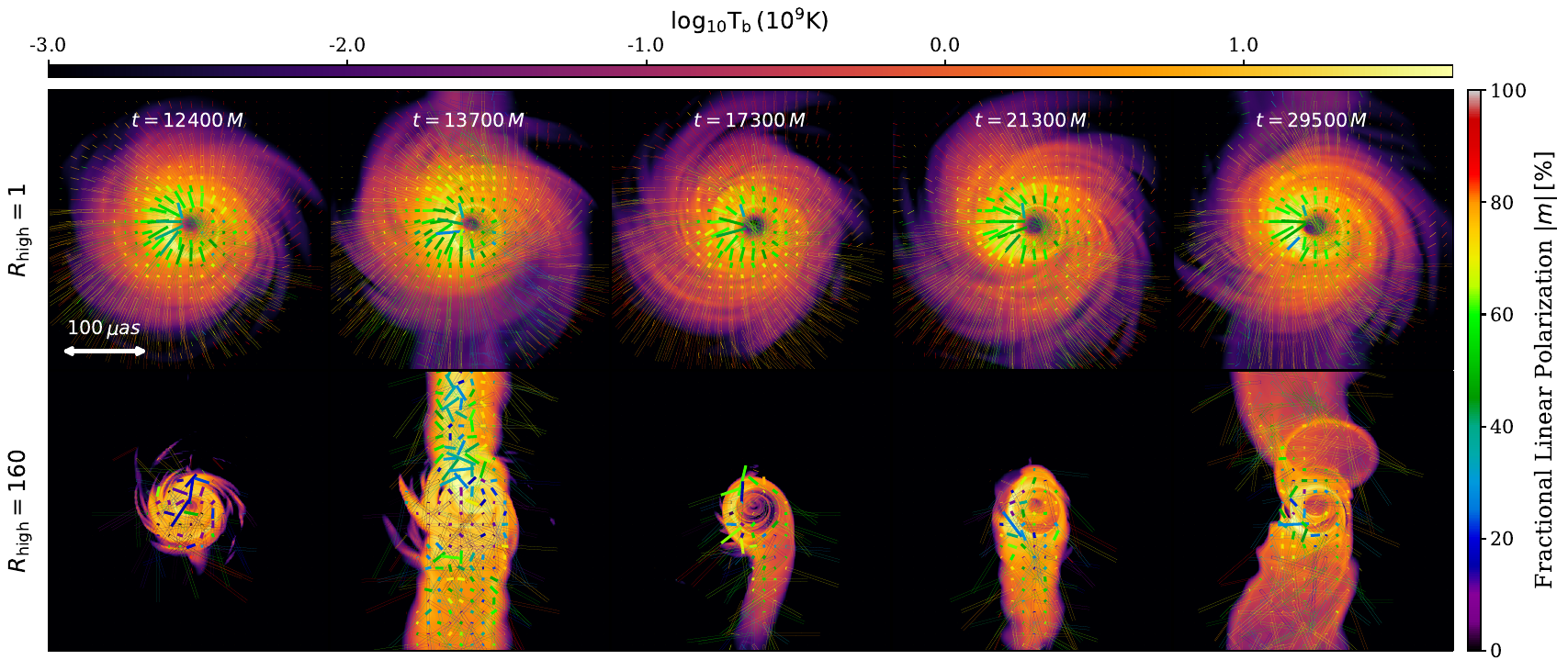}
    \caption{Ray-traced images of the logarithmic total intensity Stokes $I$ at $230 \,$GHz with $R_\text{high}=1$ ({top}) and $R_\text{high}=160$ ({bottom}). EVPAs are overlaid and colored according to their linear polarization. The images show the highly variable and transient behavior of the multi-loop model throughout the simulation.}
    \label{fig:transient_sequence}
\end{figure*}

Finally, these scalings naturally explain both the total intensity and polarization fractions shown in Fig.~\ref{fig:pol_frac}. 
For $R_{\rm high}=1$, electron heating is efficient in both low- and high-$\beta$ regions, keeping disk electrons hot (first top panel of Fig.~\ref{fig:electron_temp_rotativities}). The resulting large $\theta_e$ raises the characteristic synchrotron frequency, so that much of the disk satisfies $x=\nu/\nu_c\lesssim1$, avoiding exponential suppression. In this hot regime the emissivities scale as $j_{I,Q}\propto nB^{2/3}\theta_e^{-2/3}$ and $j_V\propto nB\theta_e^{-1}$ (Eq.~\ref{eq:emishot}), allowing the dense disk to dominate Stokes $I$, $Q$, and $V$ despite its weaker magnetic field compared to the highly magnetized but dilute jet (second top panel of Fig.~\ref{fig:electron_temp_rotativities}). Since the disk is a persistent structure in the GRMHD simulations, its emission exhibits only modest variability.
\\
Because the Stokes $Q$ emissivity has the same functional form as Stokes $I$ (Eqs.~\ref{eq:emiscold}, \ref{eq:emishot}), the linear polarization fraction broadly follows the total intensity (second panel of Fig.~\ref{fig:pol_frac}). Moreover, in the hot regime ($\theta_e\gtrsim1$), the Faraday rotation coefficient scales as $\rho_V\propto\theta_e^{-2}$ (Eq.~\ref{eq:rothigh}), resulting in weak depolarization during propagation through the disk, consistent with the low $\rho_V$ values in the third top panel of Fig.~\ref{fig:electron_temp_rotativities}. The Stokes $V$ emissivity has a steeper temperature dependence, so the circular polarization fraction is initially smaller (Eq.~\ref{eq:emishot}). Faraday conversion scales as $\rho_Q\propto\theta_e$ (Eq.~\ref{eq:rothigh}), but because the disk is only trans-relativistic ($\theta_e\lesssim1$), conversion remains weak (fourth top panel of Fig.~\ref{fig:electron_temp_rotativities}), preserving the initial ratio between linear and circular polarization throughout the simulation (second panel of Fig.~\ref{fig:pol_frac}).
In contrast, for $R_{\rm high}=160$ electron heating in the high-$\beta$ disk is inefficient, leading to much lower disk temperatures (first bottom panel of Fig.~\ref{fig:electron_temp_rotativities}). The reduced $\theta_e$ lowers $\nu_c$ and pushes most disk material into the $x\gg1$ regime, where all Stokes emissivities are exponentially suppressed (cold regime; Eq.~\ref{eq:emiscold}). The jet remains hot and strongly magnetized, satisfying $x\ll1$ and emitting efficiently (hot regime; Eq.~\ref{eq:emishot}), thereby dominating the total and polarized emission (second bottom panel of Fig.~\ref{fig:electron_temp_rotativities}). Owing to the transient nature of the jet in the multi-loop GRMHD model, this leads to strong variability and sharp drops in total intensity, particularly during jet-absent episodes (third panel of Fig.~\ref{fig:pol_frac}).
Although Stokes $Q$ should follow Stokes $I$, the linear polarization fraction is noisy and of low magnitude in the $R_{\rm high}=160$ model (bottom panel of Fig.~\ref{fig:pol_frac}). This is explained by strong Faraday rotation: linear polarization produced in the jet propagates through the cold disk, where $\rho_V\rightarrow1$ (Eq.~\ref{eq:rotlow}; third bottom panel of Fig.~\ref{fig:electron_temp_rotativities}), leading to substantial depolarization. At the same time, the hot jet exhibits strong Faraday conversion, with $\rho_Q\propto\theta_e$ (Eq.~\ref{eq:rothigh}; fourth bottom panel of Fig.~\ref{fig:electron_temp_rotativities}), efficiently converting linear into circular polarization. Together with the intrinsic jet variability, this naturally explains both the magnitude and variability of $v_{\rm net}$ in the $R_{\rm high}=160$ model.

\subsubsection{230\,GHz images}
After we report the unresolved GRRT results in previous Section we present here high-resolution images and polarization structures. In Fig. \ref{fig:transient_sequence} we present a sequence of ray-traced $230 \,$GHz images for a disk-dominated emission model ($R_{\text{high}} = 1$) in the top panels and jet-dominated one ($R_{\text{high}} = 160$) in the bottom panels at different times throughout the simulation between $t=10 \,$kM - $30\,$kM. For each image we over-plotted the EVPAs and color-coded the polarization fraction, while the length of the ticks corresponds to the normalized polarized flux (individually normalized for each snapshot). First we will discuss the total intensity, followed by the interpretation of the linear polarization structures in the next paragraph.
Our first image at t=12400\,M corresponds to a time step which is characterized by a quadrupolar magnetic field topology (see Fig.\ref{fig:parity}) and low jet power and efficiency (see Fig. \ref{fig:jet_power}). Therefore, we do not expect a see any jet contribution to the image independent of the emission model (disk or jet dominated). In contrast to the second time step at t=13700\,M. During this time the magnetic field recovered its dipolar structure (see Fig.~\ref{fig:parity}) and thus launched a powerful pair of jets, which can be seen by the increased jet efficient and jet asymmetry parameter $\mathcal{P}\sim 0$ (Fig.\ref{fig:jet_power}). This jet is clearly visible in the jet emission model (bottom panel in Fig.~\ref {fig:transient_sequence}) and indicated by extended low flux features along the north-south direction in the disk-dominated emission model (top panel in Fig.~\ref {fig:transient_sequence}). The following images correspond to time steps where the magnetic field is transitioning back to a more quadrupolar topology and to the minimum of the jet efficiency in our simulations (see Figs.~\ref{fig:parity}, ~\ref{fig:jet_power}), leading to a very faint counter-jet while the forward jet is entirely suppressed. This behavior manifests itself in the smallest emission structure throughout the sequence of images presented in Fig~\ref{fig:transient_sequence}. Shortly after this "minimum state", the accretion flow stabilizes again, i.e, the magnetic flux increases, leading to rebuilding of a dipolar magnetic field configuration in the accretion flow and thus to conditions favoring the launching of jets (Figs~\ref{fig:parity},\ref{fig:jet_power}). The associated ray-traced images show first a brightening and expansion of the counter jet and finally a fully developed bright pair of jets in the case of the jet-dominated emission model, $R_{\text{high}} = 160$. Notice that the jet signature becomes even visible in the disk-dominated emission model, $R_{\text{high}} = 1$, as faint collimated emission in the north-south direction surpasses the circular disk emission. For the influence of different choices of the $R_{\text{high}}$ parameter on the horizon scale image structure, see \citet{Jiang2024}.

\begin{figure}[h!]
    \centering
    \includegraphics[width=0.4\textwidth]{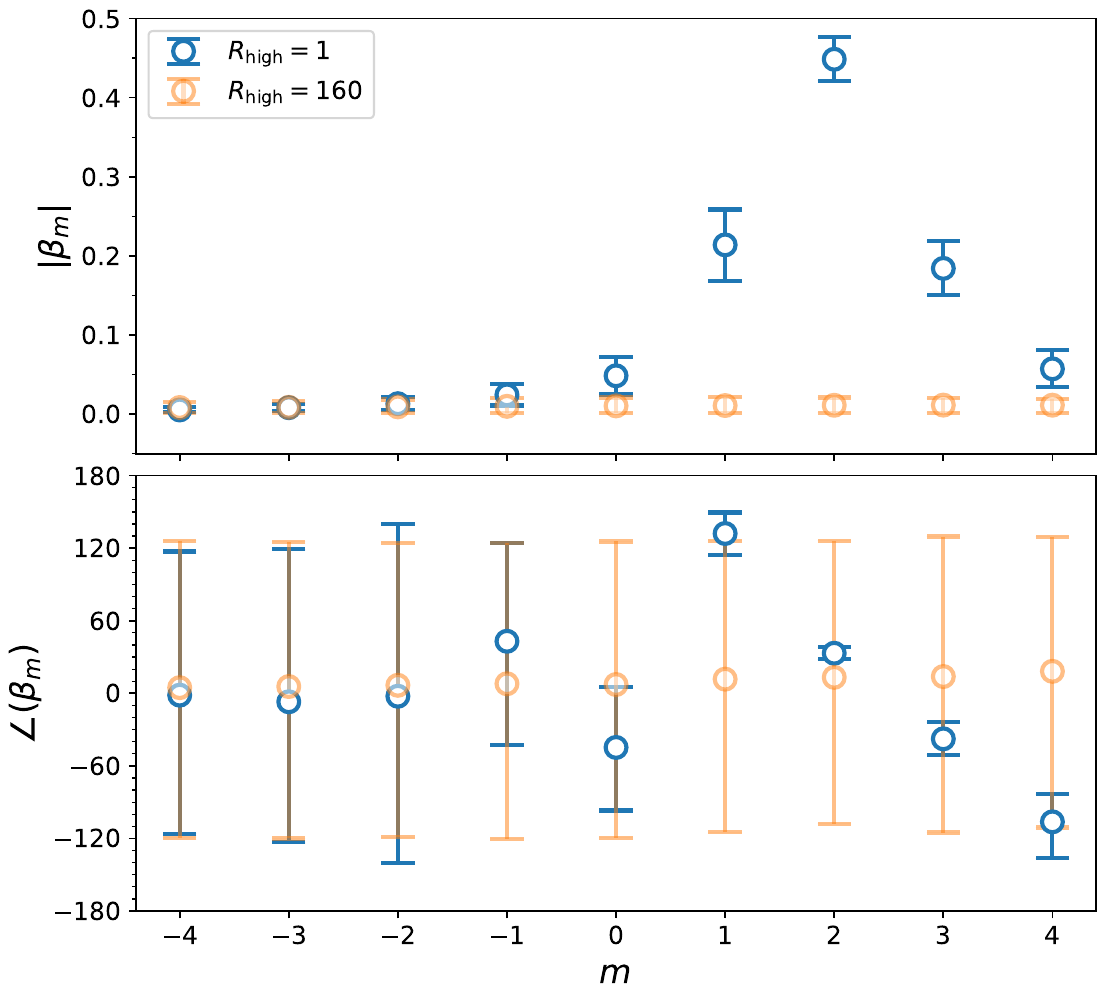}
    \caption{Fourier decomposition of the polarisation structure at 230\,GHz for $R_{\rm high}=1$ (blue) and $R_{\rm high}=160$ (orange). The top panel shows the $\left | \beta_m\right|$ and the bottom one $\angle \left(\beta_m\right)$}
    \label{fig:betamodes}
\end{figure}

To quantify the azimuthal structure of the polarized emission, we follow the Fourier decomposition of the complex linear polarization presented $P(\rho,\varphi)=Q(\rho,\varphi)+iU(\rho,\varphi)$ by \citet{Palumbo2020}. The polarization modes are defined as
\begin{equation}
\beta_m=\frac{1}{I_{\rm ann}}\int_{\rho_{\min}}^{\rho_{\max}}\int_0^{2\pi}
P(\rho,\varphi)\,e^{-im\varphi}\,\rho\,d\varphi\,d\rho ,
\end{equation}
where the normalization is given by the total annular intensity,
\begin{equation}
I_{\rm ann}=\int_{\rho_{\min}}^{\rho_{\max}}\int_0^{2\pi}
I(\rho,\varphi)\,\rho\,d\varphi\,d\rho .
\end{equation}
The amplitude $|\beta_m|$ measures the coherence of the $m$-th azimuthal polarization mode, while $\angle\beta_m$ encodes its global phase.
\\
In Fig. \ref{fig:betamodes} we perform a decomposition of the images between 10\,kM and 30\,kM for the first $\pm4$ modes. The $R_{\rm high}=1$ model exhibits enhanced low-order polarization modes, with a particularly well-defined and coherent phase in the $m=2$ component. This indicates a dominant quadrupolar polarization structure, consistent with synchrotron emission from the inner disk.
\newline In contrast, the $R_{\rm high}=160$ model shows strongly suppressed mode amplitudes and largely unconstrained phases. The cooler electron temperatures in this model increase the Faraday rotation coefficient,
$\rho_V\propto \Theta_e^{-2}$, leading to strong internal Faraday rotation that randomizes the EVPA and destroys large-scale azimuthal coherence in Stokes $Q$ and $U$. 
\newline The results of the annular Fourier decomposition is in agreement and confirm the results and discussion of the unresolved polarization signatures of the previous section.

\begin{figure*}[t]
    \centering
    \includegraphics[width=0.9\textwidth]{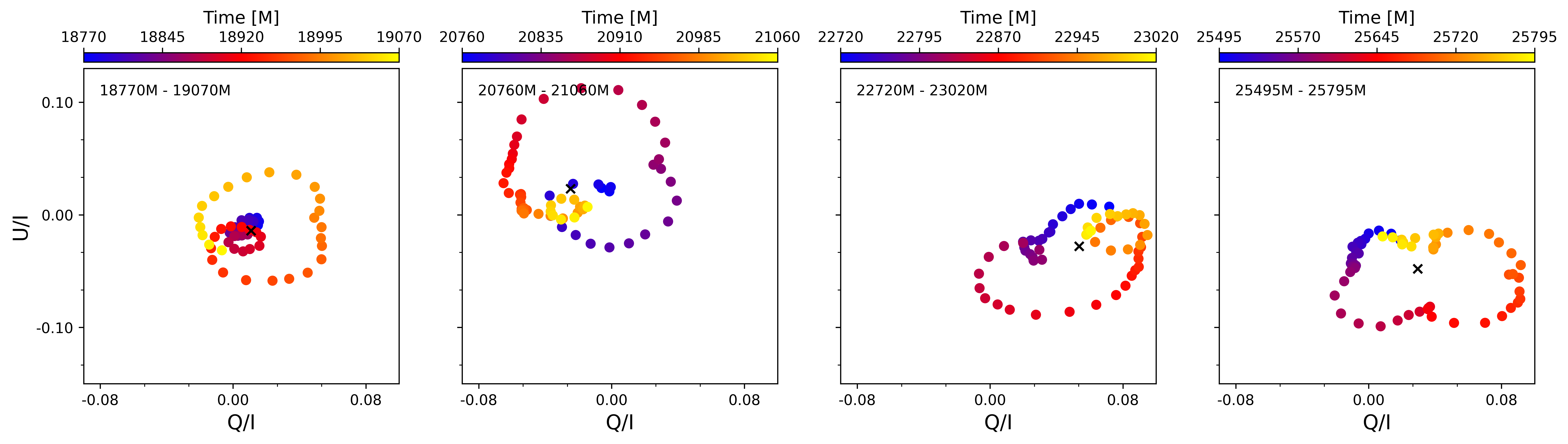}
    \caption{Coherent loops in the plane of the linearly polarized Stokes parameters $Q$ and $U$ at different times. Each loop consists of 60 points, spanning a time interval of $300\,$M.}
    \label{fig:QU_centroid}
\end{figure*}

\subsubsection{QU-Loops}
{In the introduction we mentioned that ALMA and GRAVITY observations have seen the formation of coherent loops in the $QU$-plane of the linearly polarized Stokes parameters.}
We investigate the formation of $QU$ loops following \citep{Najafi-Ziyazi2024}. The authors of that paper identify $QU$-loops during magnetic flux eruption events at times when the magnetic flux $\Phi_{BH}$ decreases sharply, while the light curve intensity rises.
Indeed, we also find that coherent $QU$-loops form in our model; however, we observe more or less pronounced loops continuously throughout the entire run-time of the simulation, rather than just during flux eruptions. Some of the most prominent loops at $230 \,$GHz and $R_\text{high} = 1$ are depicted in Fig. \ref{fig:QU_centroid}. All loops are sampled with 60 points, which corresponds to a time interval of $300 \,$M, or $102$ minutes at Sgr A*. This choice is arbitrary, but it allows a clean depiction without too many points cluttering the image, while at the same time allowing large loop structures to close themselves. The largest loop on the top right in Fig. \ref{fig:QU_centroid} has dimensions $\sim 0.12 \times 0.15$. For $R_\text{high} = 160$, the motion of the points is largely random, and we do not find strong, coherent structures in the $QU$-plane due to the depolarization caused by the Faraday effect, see Sec. \ref{sec:230GHzLightCurvesAndPolarizationFractions}. However, there is one exception of a loop, about $\sim 0.1 \times 0.1$ in size, that forms during the time interval $t=25500$- $27000 \,$M, where the linear polarization fraction dominates over the circular in the bottom panel of Fig.~\ref{fig:pol_frac}.

\subsubsection{Power spectrum densities (PSDs)}

\begin{figure}[h]
    \centering
    \includegraphics[width=0.47\textwidth]{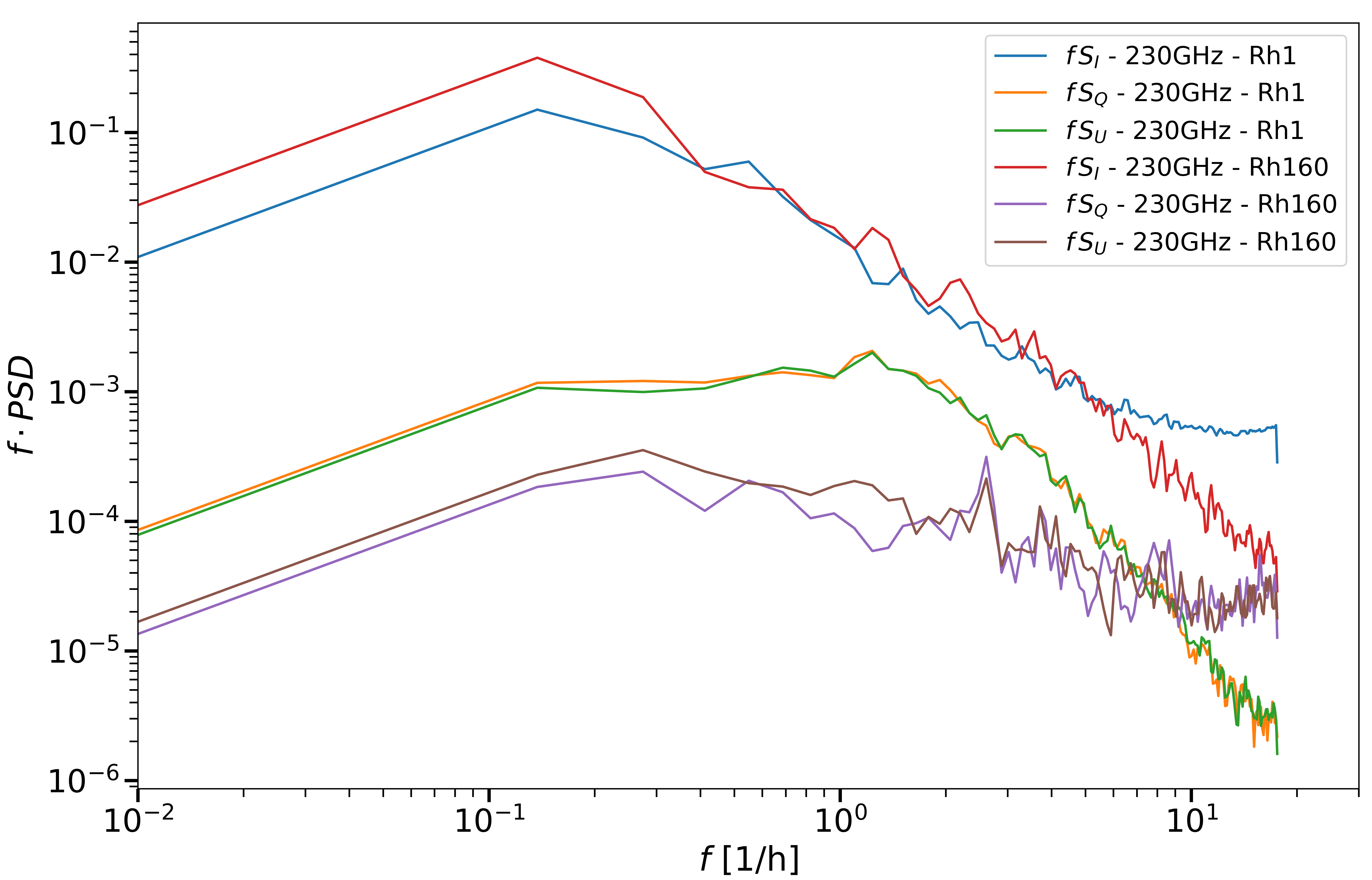}
    \caption{Power spectrum densities of Stokes $I, Q$ and $U$ for $R_\text{high}=1, 160$.}
    \label{fig:power_spectrum_densities}
\end{figure}

The power spectra of the GRRT light curves closely resemble those reported by \citet{Najafi-Ziyazi2024}, exhibiting red-noise behavior in total intensity and enhanced polarized variability at intermediate frequencies.
Stokes $I$ shows a smooth power-law decline toward higher frequencies, reflecting stochastic variability driven by turbulent accretion.
In contrast, the Stokes $Q$ and $U$ spectra display a broad excess at $f\sim0.1$--$1~{\rm h}^{-1}$, particularly in the $R_{\rm high}=1$ model, consistent with modulation by the orbital motion of coherent structures such as flux tubes or hot spots in the inner accretion flow.
As in \citet{Najafi-Ziyazi2024}, the excess is broad rather than quasi-periodic, indicating finite lifetimes and partial coherence of these structures.
\\
Analytically, coherent orbital motion at radius $r$ introduces variability near the Keplerian frequency so that emission originating from radii of a few gravitational radii naturally produces variability on timescales of tens of minutes to hours for Sgr~A$^\star$.
For a rapidly spinning black hole ($a\simeq0.94$), the orbital period at the ISCO is $T_{\rm ISCO}\sim 30~{\rm min}$, placing the expected polarized variability in the frequency range where the PSD excess is observed. The strength of this polarized excess depends sensitively on the electron temperature through Faraday effects. The Faraday rotation coefficient scales approximately as $\rho_V\propto \Theta_e^{-2}$, so hotter electrons produce substantially weaker Faraday rotation.
In the $R_{\rm high}=1$ model, the disk remains hot, resulting in low Faraday depth and allowing intrinsic polarization variability driven by orbital motion to survive radiative transfer. Conversely, in the $R_{\rm high}=160$ model, the cooler disk leads to strong, spatially varying Faraday rotation that depolarizes the emission and suppresses coherent variability in $Q$ and $U$, shifting power to lower frequencies and reducing the PSD normalization.
This combined dependence on Keplerian orbital timescales and temperature-controlled Faraday rotation naturally explains both the contrast between the two heating prescriptions and the close qualitative agreement with the polarized PSDs reported by \citet{Najafi-Ziyazi2024}.

\subsection{High Energy emission (X- and $\gamma$-rays)}
To assess high-energy signatures of multiple magnetic loops with alternating polarity, we compute Compton upscattering of low-frequency synchrotron emission using a modified version of \texttt{igrmonty}\footnote{https://github.com/afd-illinois/igrmonty} \cite{Dolence2009}. In addition to synchrotron emission, Bremsstrahlung radiation is included as a seed photon source. To ensure sufficient scattering statistics even at low inclinations ($30^\circ$), we inject $10^5$ super-photons per snapshot. We apply snapshot-dependent bias tuning to maintain an approximately unity ratio between scattered and directly emitted super-photons, which is required due to rapid variations in the fluid properties of the multi-loop accretion flow. For each GRMHD snapshot, we generate at least 10 Monte Carlo realizations to improve statistics and reduce noise \cite[see][for details]{Wong2022}. Fig.~\ref{fig:SEDs} shows the spectral energy distributions (SEDs) for four representative times (top row) and the spectral decomposition at $t=21400\,M$ (bottom row).

\begin{figure}[h]
    \centering
    \includegraphics[width=0.5\textwidth]{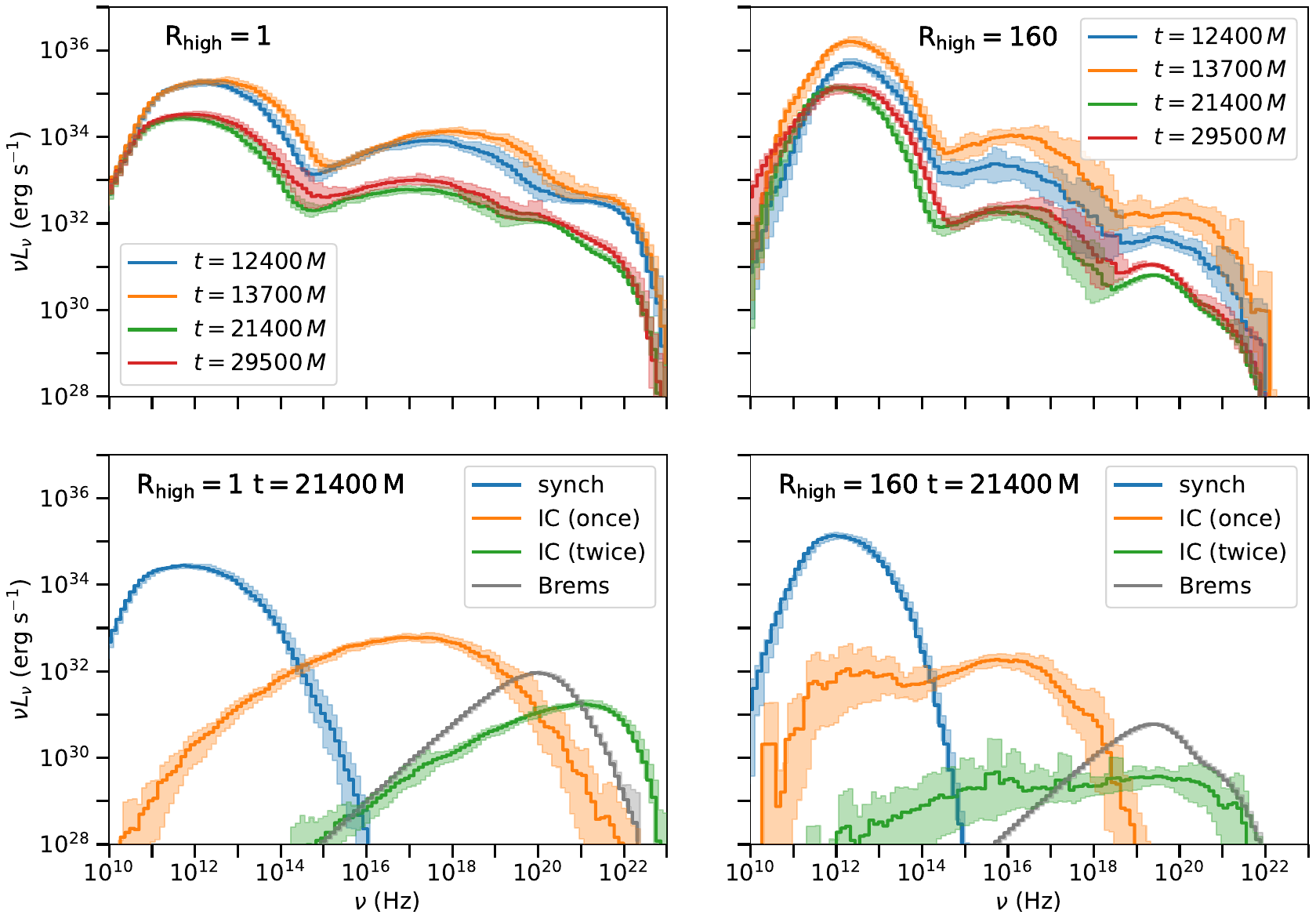}
    \caption{Spectral Energy Distribution for our models. {Top panels:} Different times steps for $R_{\rm high}=1$ (left) and $R_{\rm high}=160$ (right). {Bottom panels:}Decomposition of the SED for t=21400\,M for $R_{\rm high}=1$ (left) and $R_{\rm high}=160$ (right).}
    \label{fig:SEDs}
\end{figure}
The spectral shapes differ markedly between the two electron heating prescriptions (Fig.~\ref{fig:SEDs}). The $R_{\rm high}=1$ model exhibits broader synchrotron and inverse Compton humps and typically produces two-peaked spectra with comparatively low variability (top left panel). Spectral decomposition reveals that the hot disk dominates the emission, providing a persistent reservoir of electrons for Compton upscattering. As a result, Compton-scattered emission dominates the high-energy output despite a non-negligible Bremsstrahlung contribution (bottom left panel), yielding stable two-hump spectra.
In contrast, for $R_{\rm high}=160$ the disk electrons are significantly cooler (Fig.~\ref{fig:electron_temp_rotativities}), suppressing disk synchrotron and Compton emission. The jet therefore dominates the radiative output, but its transient nature leads to strong variability, particularly in the X-ray and $\gamma$-ray bands (Fig.~\ref{fig:highenergylc}). During jet disruptions or absences, Bremsstrahlung can temporarily dominate the high-energy emission (bottom right panel of Fig.~\ref{fig:SEDs}), producing a characteristic three-humped SED. {Fig. \ref{fig:highenergylc} shows light curves in the NIR, X-ray and $\gamma$-ray bands alongside the quiescent fluxes of the Galactic Center, as reported by GRAVITY and CHANDRA.} For $t \gtrsim 17\,\mathrm{kM}$, both models are consistent with quiescent NIR and X-ray observational constraints. {This is in contrast to standard MAD models that usually overproduce these fluxes, \citep{EHTC2022V}.}

\begin{figure}[h]
    \centering
    \includegraphics[width=0.45\textwidth]{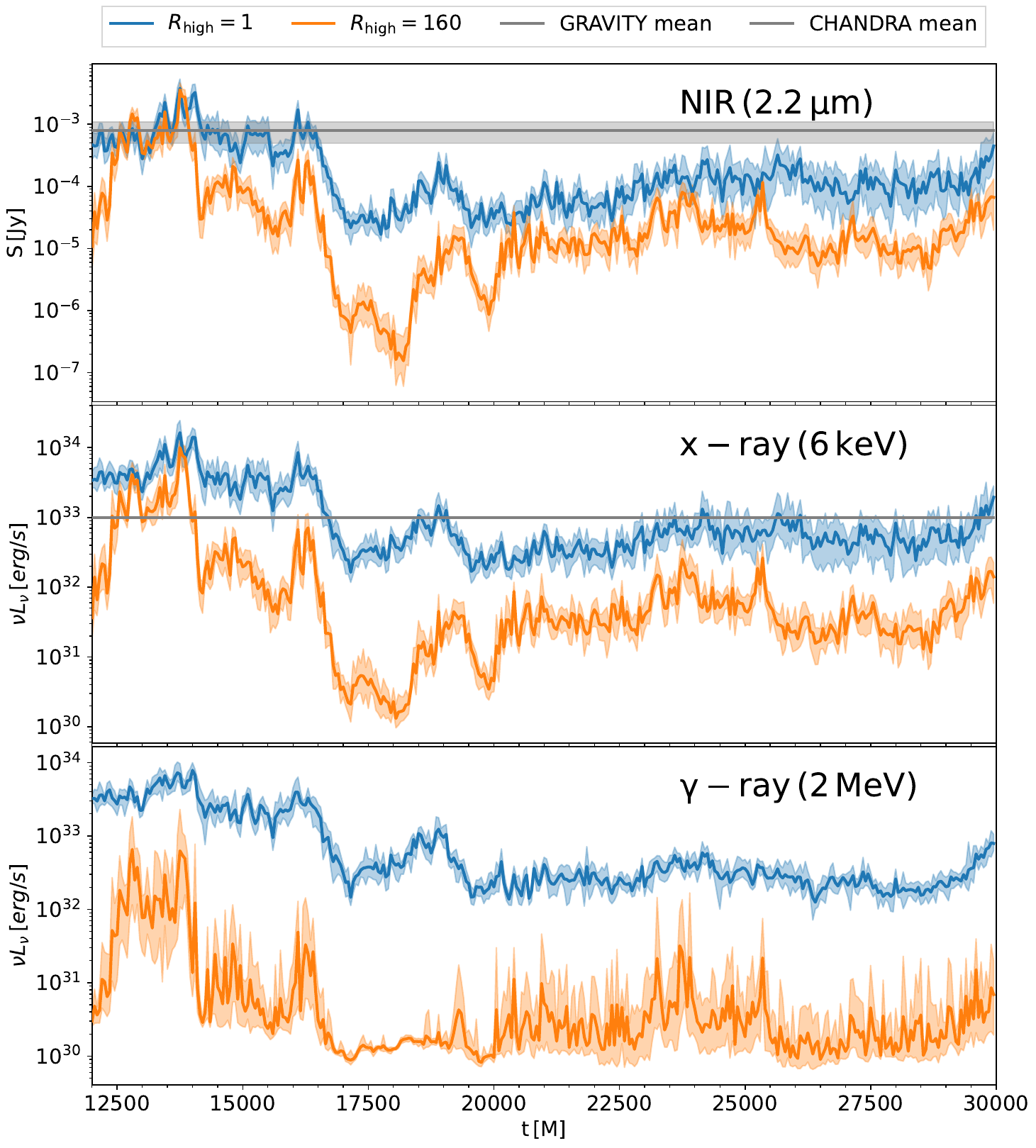}
    \caption{High-energy light curves. From top to bottom, NIR, x- and $\gamma$-rays for $R_{\rm high}=1$ and $R_{\rm high}=160$, the horizontal gray lines indicate the quiescent GRAVITY NIR flux and the upper limit on the median (quiescent) CHANDRA measurement.}
    \label{fig:highenergylc}
\end{figure}

\section{Discussion \& Conclusion}
\label{sec:discussion_conclusion}
In this work, we investigate a GRMHD simulation with multiple magnetic loops of alternating polarity and study their impact on accretion dynamics, jet formation, and radiative signatures, including millimeter images, polarization, and high-energy emission. We show that this multi-loop configuration behaves fundamentally differently from standard MAD and SANE models. Rather than individual loops being accreted sequentially, the onset of accretion distorts the initial magnetic field into large-scale kinks and turbulence near the black hole. A turbulent bubble detaches from the inner flow and propagates outward through the disk midplane, gradually mixing with the remaining ordered field. This indicates that the evolution of multi-loop accretion is governed by MRI-driven turbulence and magnetic instabilities rather than simple loop-by-loop accretion.
\\
Analysis of the magnetic flux, jet energetics, and field topology reveals that jet production is tightly coupled to the amount and symmetry of magnetic flux accumulated near the black hole. High-flux states correspond to efficient, dipole-dominated jets, whereas low-flux phases are associated with mixed or quadrupolar topologies and strongly suppressed outflows. This highly intermittent jet activity, together with rapid topology changes, clearly distinguishes the multi-loop model from canonical MAD and SANE simulations.
\\
Angular momentum transport further supports this picture. During high-efficiency jet phases, the Maxwell stress exhibits strong, coherent outward transport along the polar funnel, enabling efficient energy extraction. When the jet is quenched, the Maxwell stress weakens substantially, and turbulence dominates throughout the disk, leaving the polar regions largely devoid of outward angular momentum flux. Thus, transitions between active and inactive jet states are governed by the ability of the magnetic field to maintain large-scale coherence in the funnel.
\\
Polarized radiative transfer provides a direct observational counterpart to this dynamical behavior. At $230\,$GHz, jet features appear and disappear in close correspondence with the magnetic topology and jet efficiency. Jet-dominated models ($R_\text{high}=160$) show collimated emission during high-flux phases, while disk-dominated models ($R_\text{high}=1$) still exhibit weak north–south extensions. In contrast, during mixed or quadrupolar field configurations, both models produce compact, faint images lacking extended jet signatures. The repeated on/off switching of jet features in polarized images directly reflects the evolving magnetic structure of the multi-loop flow.
\\
For unresolved polarization fractions, only the disk-dominated ($R_\text{high}=1$) model satisfies the millimeter constraints of Sgr~A$^\star$ ($m_{\rm net}\sim7\%$, $v_{\rm net}\lesssim1\%$). As $R_\text{high}$ increases and the disk cools, the Faraday depth rises, suppressing linear polarization and enhancing circular polarization through rotation and conversion. Fig.~\ref{fig:electron_temp_rotativities} shows that $\theta_e$, $\rho_Q$, and $\rho_V$ decrease in the disk with increasing $R_\text{high}$, while remaining high in the jet. Consequently, in jet-dominated models, polarized emission undergoes strong conversion within the jet and substantial Faraday rotation while traversing the cold disk, yielding low $m_{\rm net}$ and elevated $v_{\rm net}$.
\\
These Faraday effects also shape the polarization morphology. For $R_\text{high}=1$, coherent loops in the $Q$–$U$ plane (Fig.~\ref{fig:QU_centroid}) persist throughout the simulation, with evolving sizes and orientations driven by continuous magnetic reconfigurations in the inner flow. In contrast, for $R_\text{high}=160$, strong Faraday depolarization renders the $Q$–$U$ trajectories largely stochastic, with coherent loops appearing only briefly when linear polarization dominates. This demonstrates that $QU$-loops are sensitive not only to magnetic eruptions but also to the effective Faraday depth.
\\
High-energy diagnostics provide a complementary view of the transient jet behavior. The SEDs (Fig.~\ref{fig:SEDs}) separate disk- and jet-dominated thermodynamics: the $R_\text{high}=1$ model produces a relatively stable two-hump (synchrotron + inverse Compton) spectrum, whereas the $R_\text{high}=160$ model yields a highly variable three-humped SED. When the jet is present, inverse Compton emission dominates; during jet absences, Bremsstrahlung temporarily becomes dominant, reflecting the loss of hot electrons.
\\
The high-energy light curves (Fig.~\ref{fig:highenergylc}) reinforce this distinction. The $R_\text{high}=1$ model remains near quiescent NIR and X-ray levels, while the $R_\text{high}=160$ model exhibits strong flaring tied to intermittent jet activity. This supports the interpretation that flares in low-luminosity galactic nuclei may arise from short-lived episodes of magnetic coherence that launch transient jets.
\\
Overall, our results demonstrate that multi-loop magnetic configurations naturally produce intermittent jets through self-consistent GRMHD evolution. The coupling between magnetic flux accumulation, field topology, and angular momentum transport provides a unified explanation for the rapid variability in jet power and emission. This framework offers a promising avenue for interpreting the time-variable millimeter, infrared, and high-energy emission of sources such as Sgr~A$^\star$. Future work incorporating non-thermal particles {necessary for NIR and high energy flares}, higher-resolution simulations {in order to resolve the formation of plasmoids}, as well as longer run times {to study the subsequent long-term behavior of the accretion flow and jet launching }, will further connect multi-loop magnetic dynamics to observational constraints.

\begin{acknowledgements}
This research is supported by the DFG research grant ``Jet physics on horizon scales and beyond" (Grant No.  443220636) within the DFG research unit ``Relativistic Jets in Active Galaxies" (FOR 5195). AN was supported by the Hellenic Foundation for Research and Innovation (ELIDEK) under Grant No 23698. YM is supported by the National Key R\&D Program
of China (grant no. 2023YFE0101200), the National Natural Science Foundation of China (grant no. 12273022, 12511540053), and the Shanghai municipality orientation program of basic research for international scientists (grant no. 22JC1410600).
The numerical simulations and calculations have been performed on \texttt{MISTRAL} at the Chair of Astronomy at the JMU Wuerzburg. 
\end{acknowledgements}

\bibliographystyle{aa}
\bibliography{literatur}

\appendix

\section{Numerical GRMHD}
\label{sec:GRMHD_appendix}
We solve the equations of GRMHD using the \texttt{KHARMA}\footnote{\textit{Kokkos-based High-Accuracy Relativistic Magnetohydrodynamics with Adaptive mesh refinement}} code \citep{Prather2024}. \texttt{KHARMA} is based on the \texttt{HARM}\footnote{\textit{High-Accuracy Relativistic Magnetohydrodynamics}} scheme \citep{Gammie2003, mckinney2004} and is written in C++ using the Kokkos programming model, allowing for parallelization, which makes it run efficiently on both GPUs and CPUs. For a more detailed discussion refer to the sources above.
\\
\\
In the following, greek indices $\mu, \nu$ etc. indicate spacetime coordinates and run from $0, 1, 2, 3$, where the zeroth index denotes time. Latin indices $i, j$ etc. indicate purely spatial dimensions and run from $1, 2, 3$.
\subsection{Fundamental GRMHD Equations}
The fundamental equations of GRMHD are the covariant conservation of particle number $n$, conservation of energy-momentum $T^{\mu}\,_\nu$ and the homogeneous Faraday's law, respectively
\begin{equation}
\label{covariantConservationLaws}
\begin{aligned}
    \nabla_\mu (n\, u^\nu) = 0, \\
    \nabla_\mu T^\mu\,_\nu = 0, \\
    \nabla_\mu \,^*F^{\mu \nu} = 0.
\end{aligned}
\end{equation}
Rewriting these equations in a coordinate basis, the first equation becomes
\begin{equation}
\label{numberConservation}
    \frac{1}{\sqrt{-g}} \partial_\mu \big( \sqrt{-g} \rho u^\mu \big) = 0,
\end{equation}
where $\rho$ is the rest mass density, $u^\mu$ is the 4-velocity of the fluid and $g = det(g_{\mu \nu})$ is the determinant of the metric tensor.\\
Next, the conservation of energy-momentum, written in a coordinate basis is
\begin{equation}
\label{basisEnergyConservation}
    \partial_\mu \big( \sqrt{-g} \, T^t\,_\nu \big) = - \partial_i \big( \sqrt{-g} \, T^i\,_\nu \big) + \sqrt{-g} \, T^\kappa \,_\lambda \Gamma^\lambda\,_{\nu \kappa},
\end{equation}
where $T^\mu \, _\nu$ is the stress-energy-momentum tensor and $\Gamma^\lambda \,_{\nu \kappa}$ are the connection coefficients. For a magnetized plasma the energy-momentum tensor is made up of a perfect fluid part and a part for the electromagnetic field.\\
To see this let us consider Ohm's law, which is given in the MHD approximation by
\begin{equation}
\label{MHDOhmsLaw}
    \boldsymbol{E} + \boldsymbol{u} \times \boldsymbol{B} = \eta \boldsymbol{J}
\end{equation}
In the limit of the ideal MHD approximation without resistivity ($\eta \approx 0$) the plasma is a perfect conductor and the right-hand side of Eq. (\ref{MHDOhmsLaw}) vanishes. Noticing that the left-hand side is equal to the electric field $\boldsymbol{E'}$ when boosting into the fluid frame, we have
\begin{equation}
\label{idealMHDOhmsLaw}
    \boldsymbol{E'} = \boldsymbol{E} + \boldsymbol{u} \times \boldsymbol{B} = \boldsymbol{0},
\end{equation}
meaning the electric field vanishes in the fluid rest frame due to the high conductivity of the plasma, i.e. the Lorentz force vanishes in the fluid frame. This can be expressed in a covariant way using the Faraday tensor $F^{\mu \nu}$ as
\begin{equation}
\label{covariantLorentzForceIdealMHD}
    u_\mu F^{\mu \nu} = 0.
\end{equation}
Using the dual Faraday tensor one can define the magnetic field 3-vector $B^i = \, ^*F^{it}$, \citep{Komissarov1999}. And the the components of the magnetic field 4-vector are related to $B^i$ by
\begin{equation}
\label{b^t}
    b^t = g_{i \mu} B^i u^\mu
\end{equation}
and
\begin{equation}
\label{b^i}
    b^i = \frac{B^i + b^t u^i}{u^t}
\end{equation}
The MHD energy-momentum tensor takes the form
\begin{equation}
\label{MHDenergyMomentumTensor}
    T^{\mu\nu} = (\rho + u + p + b^2) u^\mu u^\nu + (p + \frac{b^2}{2})g^{\mu \nu} - b^\mu b^\nu,
\end{equation}
where $\rho$ is the rest mass density, $p$ is the pressure, $u$ is the internal energy.\\
Finally, the induction law ($\nabla \times \boldsymbol{E} = -\frac{\partial \boldsymbol{B}}{\partial t}$) and the condition of a divergence-free magnetic field ($\nabla \boldsymbol{B} = 0$) in Maxwell's equations can be written in the form
\begin{equation}
\label{relativisticInduction}
    \partial_t \big( \sqrt{-g} \, B^i \big) = -\partial_j \big[ \sqrt{-g} \, (b^j u^i - b^i u^j)  \big]
\end{equation}
and
\begin{equation}
\label{noMonopoles}
    \frac{1}{\sqrt{-g}} \partial_i \big( \sqrt{-g} \, B^i\big) = 0.
\end{equation}
In summary, the code evolves a magnetized fluid by solving the system of equations given by particle number conservation Eq. (\ref{numberConservation}), the conservation of the energy-momentum in a coordinate basis Eq. (\ref{basisEnergyConservation}), using the ideal MHD energy-momentum tensor from Eq. (\ref{MHDenergyMomentumTensor}), as well as the induction equation Eq. (\ref{relativisticInduction}), respecting the constraint of a divergence-free magnetic field Eq. (\ref{noMonopoles}).\\
To close the system of conservation laws we assume an equation of state for an ideal gas, which allows us to relate the specific enthalpy $h$ to the density $\rho$ and gas pressure $P$, given some adiabatic index $\gamma$:
\begin{equation}
\label{noMonopoles}
    h = 1 + \frac{\gamma P}{(\gamma - 1) \rho}.
\end{equation}

\subsection{Numerical Scheme}
The numerical scheme employed by \texttt{HARM} is conservative, meaning there is a set of physical quantities that are conserved throughout the evolution of the simulation. In general, if some quantity $Q$ is conserved it obeys a continuity equation of the form
\begin{equation}
\label{continuityEquation}
    \frac{\partial Q}{\partial t} + \nabla \boldsymbol{F}(Q) = 0,
\end{equation}
where $\boldsymbol{F}(Q)$ denotes the associated flux of the quantity $Q$. These fluxes are the means by which the quantities are updated within the cells.\\
For numerical computations it is more convenient to convert the differential form of Eq. (\ref{continuityEquation}) to an integral formulation, and employ a finite-volume method to represent and evaluate the partial differential equations.\\
To this end, we integrate Eq. (\ref{continuityEquation}) over the small cell volume $\Delta V$ and use the divergence theorem to rewrite the volume integral in terms of an integral over the surface $\boldsymbol{S}$ enclosed by that cell volume and obtain

\begin{equation}
\label{finiteVolumeMethod}
\begin{split}
    \frac{d\bar{Q}}{dt} + \frac{1}{\Delta V} \oint\limits_{\partial (\Delta V)} \boldsymbol{F}(Q) \, d\boldsymbol{S} = 0.
\end{split}
\end{equation}
with $\bar{Q} = \frac{1}{\Delta V}\int\limits_{\Delta V} Q \, dV$ is the average value of $Q$ over the cell volume. Finally, we can write out the surface integral in Eq. (\ref{finiteVolumeMethod}) explicitly in terms of the fluxes entering and leaving through the faces of the cell $(i, j, k)$, see, e.g. \citep{Porth2017} for a detailed discussion
\begin{equation}
\label{finiteVolumeMethod2}
\begin{split}
    \frac{d\bar{Q}_{i, j, k}}{dt} = - \frac{1}{\Delta V} &\bigg[\boldsymbol{F}^1\,\Delta S^1|_{i+1/2, j, k} -\boldsymbol{F}^1\,\Delta S^1|_{i-1/2, j, k} \\ 
    & + \boldsymbol{F}^2\,\Delta S^2|_{i, j+1/2, k} - \boldsymbol{F}^2\,\Delta S^2|_{i, j-1/2, k} \\& + \boldsymbol{F}^3\,\Delta S^3|_{i, j, k+1/2} - \boldsymbol{F}^3\,\Delta S^3|_{i, j, k-1/2}\bigg].
\end{split}
\end{equation}
In GRMHD codes, there are typically a number of physical quantities that are conserved, and they are encompassed in a vector $\boldsymbol{U}$ and a flux tensor $\boldsymbol{F}$ and Eq. (\ref{continuityEquation}) holds for each component.\\
In the case of \texttt{HARM} the vector of conserved variables is given by
\begin{equation}
\label{conservedVariables}
    \boldsymbol{U} = \sqrt{-g} \big( \rho u^t , T^t_t, T^t_i, B^i \big),
\end{equation}
and the primitive variables are
\begin{equation}
\label{primitiveVariables}
    \boldsymbol{P} = \big( \rho, u, v^i, B^i\big),
\end{equation}
where $v^i = u^i/u^t$ are the components of the 3-velocity.
\\
In order to find the conserved variables $\boldsymbol{U}(\boldsymbol{P})$ and fluxes $\boldsymbol{F}(\boldsymbol{P})$ as functions of the primitive variables, one must find $u^t$ by solving the relativistically invariant equation $g_{\mu \nu} u^\mu u^\nu = u^\nu u_\nu = -1$. We can find $b^\mu$ with the help of $v^i$ and $B^i$ and Eqs. (\ref{b^t}) and (\ref{b^i}) and finally substitute into the equation for the energy-momentum tensor Eq. (\ref{MHDenergyMomentumTensor}).\\
In order to update the conserved variables, one must perform the inversion and solve for $\boldsymbol{P}(\boldsymbol{U})$ at the end of each time step.

\section{Conversion to CGS Units}
\label{CGSconversion}
Since the GRMHD simulation is scale-free, the quantities we obtain in the simulation data are unit-less and we have to rely on observational parameters such as the black hole mass, distance and accretion rate. Since the mass of the black hole and its distance are well known, we can use them directly in our radiative transfer calculations. We use a black hole mass of $M=4.14\times10^6 \, M_\odot$ and distance $d = 8.14 \,$kpc to Sgr~A$^\star$ as reported by the \cite{Gravity2022}. With that the simulation length scale $L_\text{unit} = r_g = GM/c^2$ and the time unit is $T_\text{unit} = r_g/c = GM/c^3$ are fixed. However, the mass accretion rate of a particular system is in general not well constrained and therefore becomes a free parameter $M_\text{unit}$. We therefore use the flux as another well constrained observable and vary the value of the accretion rate $M_\text{unit}$ such that the resulting average flux of the GRRT calculation matches the observed value of some object of interest. In our case this is $2.4\,$Jy from $230\,$GHz observations of Sgr~A$^\star$ by the EHT, \citep{EHTC2022I}. The same $M_\text{unit}$ can then also be used for GRRT calculations at different frequencies. For $R_\text{high} = 1$ we found $M_\text{unit} = 2.683 \times 10^{17}$g. For $R_\text{high} = 160$ we used $M_\text{unit} = 7.796 \times 10^{18}g$.\\
With $M_\text{unit}$ fixed we now have access to a whole range of parameters, since it defines the density unit of the system $\rho_\text{unit} = M_\text{unit}/L_\text{unit}^3$. We can use this to calculate the units for quantities such as the internal energy $u_\text{unit} = \rho_\text{unit}c^2$ and the magnetic field $B_\text{unit} = c\sqrt{4\pi\rho_\text{unit}}$ and move to physical CGS units by multiplying the desired quantity by its respective unit value. We compiled a list of these conversion factors in Tab. \ref{tab:unit_conversion_table}.

\begin{table}[t]
\begin{tabular}{lccc}
\textrm{~} & $R_\text{high}=1$ & $R_\text{high}=160$  \\
\hline\hline
$L_\text{unit}$ [cm]  & $6.11\cdot10^{11}$  & $6.11\cdot10^{11}$  \\
$T_\text{unit}$ [s]  & $20.40$  & $20.40$  \\
$M_\text{unit}$ [g]  & $2.683\,\cdot\,10^{17}$  & $7.796\,\cdot\,10^{18}$  \\
$\rho_\text{unit}$ [g cm$^{-3}$] & $1.17\,\cdot\,10^{-18}$  & $3.41\,\cdot\,10^{-17}$ \\
$B_\text{unit}$ [G]  & $115.12$  & $620.54$  \\
\hline
\end{tabular}
\caption{Unit quantities for conversion to CGS units.}
\label{tab:unit_conversion_table}
\end{table}

\section{Comparison to MAD model}
\label{MADcomp}

\begin{figure}
    \centering
    \includegraphics[width=.5\textwidth]{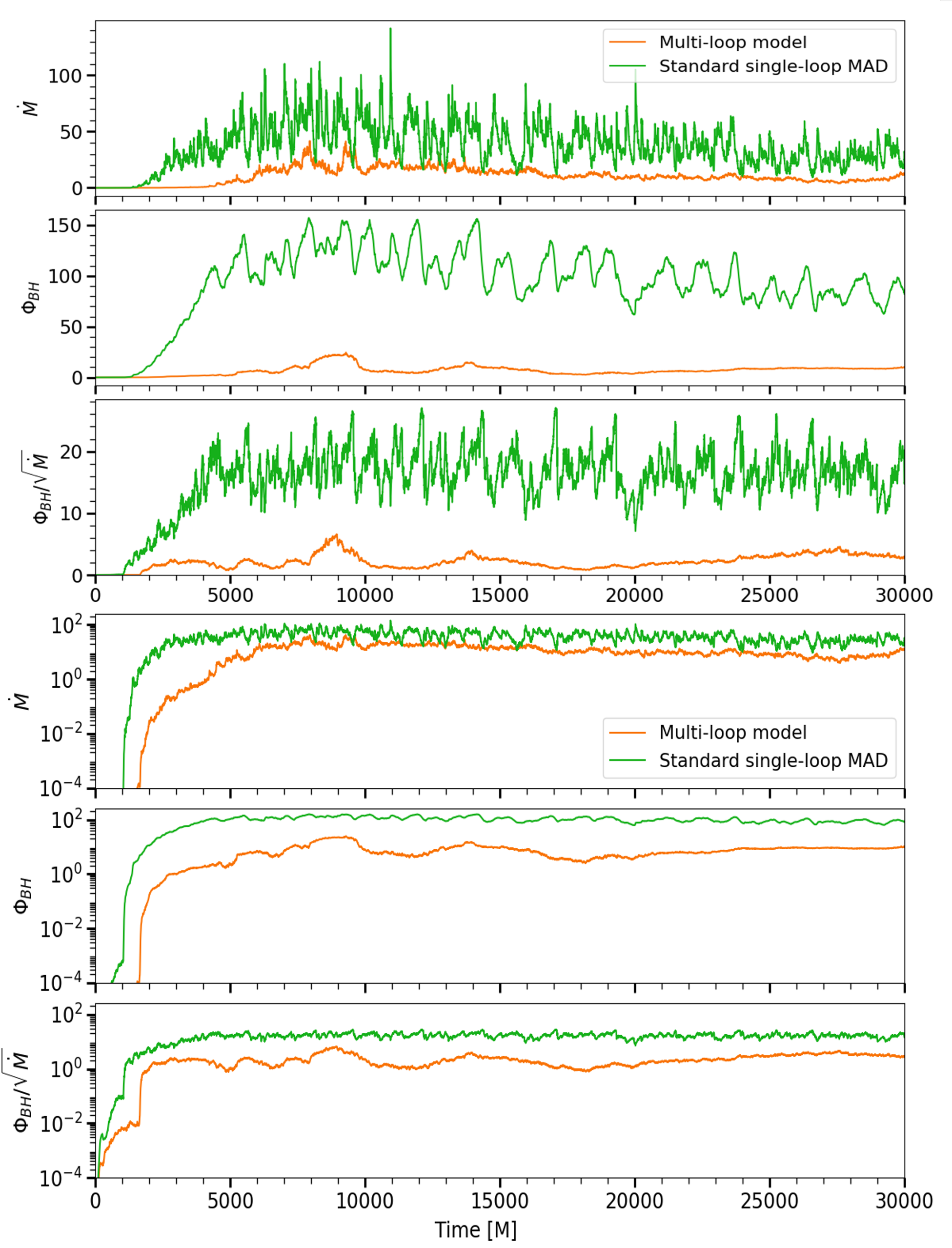}
    \caption{MAD vs. Mloop}
    \label{fig:MADcomp}
\end{figure}

\end{document}